\documentclass[aps,pra,twocolumn,amsmath,amssymb,footinbib,showpacs,longbibliography,superscriptaddress]{revtex4-1}

\newcommand{\Jnature}{Nature (London)}
\newcommand{\Jnatphys}{Nat. Phys.}

\newcommand{\Jprl}{Phys. Rev. Lett.}

\newcommand{\Jprb}{Phys. Rev. B}

\newcommand{\Jrmp}{Rev. Mod. Phys.}

\newcommand{\Jepl}{Europhys. Lett.}

\newcommand{\Jannphys}{Ann. Phys. (NY)}

\usepackage{times}
\usepackage[english]{babel}
\usepackage{latexsym}
\usepackage{graphics}
\usepackage{subfigure}
\usepackage{epsfig}
\usepackage{color}
\usepackage{hyperref}
\usepackage{braket} 
\usepackage{amstext}
\usepackage{amssymb}
\usepackage{graphicx}
\usepackage{esint}
\usepackage{braket}
\usepackage{babel}
\usepackage{amsmath}
\hypersetup{
colorlinks=true,
citecolor=blue,
linkcolor=red,
urlcolor=black
}


\begin{document}

\title{Dynamics of the Bose-Hubbard Model Induced by On-Site or Long-Range Two-Body Losses}

\author{Julien Despres}
\affiliation{JEIP, USR 3573 CNRS, Coll\`ege de France, PSL Research University, 11 Place Marcelin Berthelot, 75321 Paris Cedex 05, France}

\author{Leonardo Mazza}
\affiliation{Universit\'e Paris-Saclay, CNRS, LPTMS, 91405, Orsay, France}
\affiliation{Institut Universitaire de France, 75005 Paris, France}

\author{Marco Schir\`o}
\affiliation{JEIP, USR 3573 CNRS, Coll\`ege de France, PSL Research University, 11 Place Marcelin Berthelot, 75321 Paris Cedex 05, France}

\date{\today}

\begin{abstract}
We present a theoretical study of the dissipative dynamics of the Bose-Hubbard model induced by on-site or long-range two-body losses. 
We first consider the one-dimensional chain and the two-dimensional square lattice, and study the dynamics induced by the sudden switch-on of two-body losses on a weakly-interacting superfluid state. 
The time-dependent density is obtained in the spirit of the Bogolyubov approach by calculating theoretically the equations of motion associated to the relevant quadratic bosonic correlators. 
In the one-dimensional case, our results compare very well with quasi-exact numerical calculations based on the quantum jump method implemented using tensor networks. 
We find that the intermediate-time dynamics of the density displays an algebraic decay characterized by an interaction-dependent power-law exponent. The latter property still holds for long-range two-body loss processes but it is absent in the two-dimensional square lattice with on-site losses. 
\end{abstract}

\maketitle

\section{Introduction}
\label{sec:intro}

In the last decades, the decisive advances in the experimental control of quantum matter have given consequent momentum to the experimental and theoretical 
investigation of the quench dynamics of isolated quantum lattice models~\cite{jurcevic2014,richerme2014,cheneau2012,polkovnikov2011,gogolin2016,calabrese2005}. However, since quantum systems cannot be perfectly isolated from surrounding environments, the study of open quantum systems has attracted a lot of interest~\cite{zoller2008,buchler2008,cirac2009} and has focused on phenomena such as dephasing noise~\cite{esposito2005,gaspard2005,Eisler_2011,kollath2018,turkeshi2021diffusion}, incoherent hopping~\cite{alba2020} and  gain/loss processes~\cite{langen2016,rauer2016,syassen2008,bo2013,ott2013,krapivsky2019,krapivsky2020,sels2020,alba2022,seetharam2022correlation,marino2022universality, rosso2021, rosso2023,mazza2023dissipative}. 
Under the approximation of a Markovian environment and of weak system-bath coupling,
these open quantum many-body systems are described by the well-known Lindblad master equation~\cite{fazio2024,breuer2007,ashida2020,daley2014} for the time evolution of the system density matrix, 
which displays a unitary part and a dissipative part, the former written in terms of a Hamiltonian and the latter in terms of the so-called Lindblad jump operators. 

Experimentally, quantum simulators based on trapped ultracold atoms loaded in an optical lattice generated by the interference of counter-propagating laser beams permit to simulate quantum lattice models~\cite{bloch2012,gross2017}, such as the Bose-Hubbard model~\cite{bakr2010,chen2011}. 
Its out-of-equilibrium properties are accessible via quantum quenches realized experimentally by suddenly modifying the intensity of the laser beams controlling the artificial lattice depth and thus the ratio between the hopping amplitude and the two-body repulsive interaction strength~\cite{cheneau2012,trotzky2012,fischer2006,fischer2008}. 
Loss processes, inducing a rich  physics including the quantum Zeno effect~\cite{syassen2008,ripoll2009,schiro2022}, the loss-assisted quantum control~\cite{dong2022} or the loss-induced cooling effect~\cite{kohl2024}, can be studied using this experimental platform, as losses appear naturally in these experiments. 
Indeed, one-body losses result from scattering with background or thermal atoms~\cite{knoop2012}, two-body losses can be engineered 
by light-assisted inelastic two-body collisions which can also occur naturally in ultracold-atom experiments~\cite{aspect2000,franchi2017,tomita2017,syassen2008,weiner1999,bouganne2020} and three-body losses, where a highly bound diatomic molecule is formed, are always present and dominate in general the global loss process~\cite{soding1999,tolra2004,weber2003}.  
Most importantly for this work, in several experimental platforms, two-body losses can be engineered with light-induced collisions and turned on or off at will~\cite{tomita2017,syassen2008,weiner1999,bouganne2020}.

In this work, we investigate the dynamics of the dissipative Bose-Hubbard model driven out of equilibrium by suddenly turning on on-site or long-range two-body losses (the term \textit{on-site} 
refers to a loss process involving a single lattice site whereas the term \textit{long-range} implies also
distinct lattice sites) where the latter process arises from long-range dipole-dipole interactions in ultracold gases of polar molecules~\cite{ni2010}.
The quench dynamics associated to the dissipative Bose-Hubbard model induced by many-body losses has been the object of recent analytical and numerical works~\cite{johnson2017longlived,bouchoule2020theeffect,bouchoule2021losses,liu2022,liu2024,Rowlands_2024,nagao2024} hence, when possible, we will compare our results with those provided by the previous references. 

We first introduce a theoretical approach based on the Bogolyubov description of weakly-interacting Bose gases and derive the equations of motion associated to the relevant quadratic bosonic correlators for both on-site and long-range two-body losses which are valid when the initial state is in the superfluid-mean-field regime. For the one-dimensional chain and various interaction and dissipation strengths, we benchmark our theoretical predictions with numerical results obtained via the quantum jump method implemented using tensor networks. 
Focusing on the intermediate-time dynamics of the density, an algebraic decay in time characterized by an interaction-dependent power-law exponent is found. 
Finally, for the two-dimensional Bose-Hubbard model with on-site two-body losses we find that the time evolution of the density is interaction- and dissipation-independent and remains very similar to the non-interacting case. This dependence on the lattice dimensionality of the density profile in time is explained using simple physical arguments supported by analytical ones. 

The paper is organized as follows: in Sec. \ref{sec:model}, we start by introducing the model and the quench protocol. In Sec.~\ref{sec:methods}, we introduce the theoretical and numerical methods used throughout the paper. In Sec.~\ref{sec:BH_chain}, we discuss the dissipative quench dynamics of the Bose-Hubbard chain in the framework of on-site two-body losses. In Sec.~\ref{sec:non_local}, we move on to the case of long-range two-body losses. Then, in Sec.~\ref{sec:2D}, we investigate theoretically the dissipative quench dynamics of the two-dimensional Bose-Hubbard model on a square lattice for on-site two-body losses. Finally in Sec.~\ref{sec:conclusion}, we present our conclusions. In the Appendices we include further technical details regarding our theoretical and numerical results.

\section{Model and quench protocol}
\label{sec:model}

We consider the one-dimensional Bose-Hubbard model (1D BH) on a lattice of length $L$ whose lattice spacing is fixed to unity
($a = 1$) with periodic boundary conditions; for simplicity, we also set $\hbar = 1$. The corresponding Hamiltonian $\hat{H}$ reads:
\begin{equation}
\label{bhm}
\hat{H} = -J \sum_{R} \left(\hat{b}^{\dag}_{R} \hat{b}_{R+1} + \mathrm{h.c.}\right)+\frac{U}{2}\sum_R\hat{n}_{R}(\hat{n}_{R}-1),
\end{equation}
where $\hat{b}_R$ and $\hat{b}_{R}^{\dag}$ denote the bosonic annihilation and creation operators acting on the lattice site $R \in \mathbb{N}$,
$\hat{n}_R = \hat{b}^{\dag}_R \hat{b}_R$ refers to the local occupation number, $J > 0$ corresponds to the hopping amplitude and $U>0$ is the on-site repulsive interaction strength. 
At zero-temperature, the phase diagram of the BH chain has been extensively studied~\cite{sachdev2001,cazalilla2011}; it displays a gapless superfluid (SF) and a gapped Mott-insulating (MI) phase, determined by the competition between the hopping term, the repulsive interactions and the average filling $\bar{n}$. For $\bar{n} \in \mathbb{N}^*$, the SF-MI phase transition is of the Berezinskii-Kosterlitz-Thouless type and for $\bar{n} = 1$ the critical value is $(U/J)_{\mathrm{c}} \simeq 3.3$~\cite{kuhner2000,kashurnikov1996exact,ejima2011,rombouts2006}. For non-integer fillings, the quantum system remains in the SF phase for any value of the dimensionless interaction parameter $U/J$. 

The dissipative quench dynamics associated to the 1D BH model is fully characterized by the Lindblad master equation:
\begin{align}
\frac{\mathrm{d}}{\mathrm{d}t}\hat{\rho}(t) = -i \big[\hat{H},\hat{\rho}(t)\big] + \sum_{\lambda} \hat{L}_{\lambda} \hat{\rho}(t) \hat{L}^{\dag}_{\lambda} - \frac{1}{2} \big \{ \hat{L}^{\dag}_{\lambda} \hat{L}_{\lambda}, \hat{\rho}(t) \big \}
\label{lind_mast_eq}
\end{align}
where $\hat{H}$ is the Hamiltonian of the BH chain defined at Eq.~\eqref{bhm}. 
$\hat{\rho}$ is the time-dependent density matrix and $\hat{L}_{\lambda}$ the Lindblad jump operator acting on the different degrees of freedom contained in
$\lambda$. For on-site two-body losses, $\lambda = \{R\}$ and
\begin{align}
 \hat{L}_R = \sqrt{\gamma} \hat{b}_R^2
\end{align}
  where $\gamma$ corresponds to the local dissipation rate. For long-range two-body losses, $\lambda = \{R,R'\}$ contains two lattice site indices and the jump operator reads
 \begin{align}
 \hat{L}_{R,R'} = \sqrt{\gamma_{|R-R'|}} \hat{b}_R \hat{b}_{R'}, \gamma_{|R-R'|} = \frac{\Gamma}{(1+|R-R'|)^{\alpha}}.
 \end{align} 
The long-range dissipation strength $\gamma_{|R-R'|}$ has thus an algebraic decay defined by $\Gamma$ the on-site dissipation rate and $\alpha$ the power-law exponent governing the algebraic decay in real space. Note that the analytical expression of $\gamma_{|R-R'|}$
preserves translational invariance and thus allows for a description of losses in reciprocal space. 

To drive the BH chain out of equilibrium, the following quench protocol is considered. We start from an initial many-body quantum state corresponding to the ground state of the BH chain without dissipation implying $\gamma = 0$ ($\Gamma = 0$) for on-site (long-range) two-body loss processes; the initial density (or equivalently the filling $\bar n$ since $a = 1$ is unitary) is ${n} = 1$ and we will consider the specific case of a SF. At time $t=0$ we let the system evolve with a non-zero value of the dissipation strength $\gamma > 0$ ($\Gamma > 0$) without changing the Hamiltonian parameters. 

\section{Theoretical and numerical methods}
\label{sec:methods}
In this section we present the theoretical and numerical methods employed in this paper. 
\subsection{Bogolyubov theory and mean-field decoupling}

We focus here on the theoretical methods for the dissipative quench dynamics induced by on-site two-body losses of the 1D BH model initially confined in the (weakly-interacting) SF-mean-field regime.
This specific regime within the SF phase is characterized by a small dimensionless interaction parameter with respect to the filling, i.e. $U/J \ll \bar{n}$, see Ref.~\cite{despres2019}. 

Following a standard approach, the Hamiltonian $\hat{H}$ at Eq.~\eqref{bhm} can be expressed as a quadratic bosonic form in momentum space and then diagonalized using a bosonic Bogolyubov transformation. In what follows, 
we briefly review the main ideas. Relying on the mean-field approximation stating that the $k = 0$ mode is macroscopically occupied, the 1D BH Hamiltonian $\hat{H}$ takes the following form:
\begin{equation}
\hat{H} = \frac{1}{2} \sum_{k \neq 0}\mathcal{A}_{k}\left(\hat{b}^{\dag}_{k} \hat{b}_{k}+\hat{b}_{-k}\hat{b}^{\dag}_{-k}\right)+\mathcal{B}_{k} \left(\hat{b}^{\dag}_{k} \hat{b}^{\dag}_{-k}+\hat{b}_{k} \hat{b}_{-k} \right)
\label{eq_bhm_quadratic}
\end{equation}
where the coefficients are given by $\mathcal{A}_{k} = 4J\sin^2(k/2) + U\bar{n}$ and $\mathcal{B}_{k} = U \bar{n}$~\cite{roux2013,despres2019}. 
We then rely on the following Bogolyubov transformation:
$\hat{b}_{k} = u_{k} \hat{\beta}_{k} + v_{k} \hat{\beta}^{\dag}_{-k},$ with $u_k$ and $v_k$ real functions. $\hat{\beta}_k$ and $\hat{\beta}^{\dag}_k$ denote the bosonic Bogolyubov annihilation and creation operators acting on the quasi-momentum $k$ and obeying canonical commutation rules namely $[\hat{\beta}_{k}, \hat{\beta}^{\dag}_{k'}] = \delta_{k,k'}$ and $[\hat{\beta}_{k},\hat{\beta}_{k'}] = [\hat{\beta}^{\dag}_{k}, \hat{\beta}^{\dag}_{k'}] = 0$. This implies $u_{k}^{2} - v_{k}^{2} = 1$, and by choosing
\begin{equation}
u_{k} = \left[ \frac{1}{2} \left( \frac{\mathcal{A}_{k}}{\mathcal{E}_{k}} + 1 \right) \right]^{1/2},v_k = -\left[ \frac{1}{2} \left( \frac{\mathcal{A}_{k}}{\mathcal{E}_{k}} - 1 \right) \right]^{1/2}
\end{equation}
$\hat H$ takes a diagonal form and the corresponding quasiparticle dispersion relation $\mathcal{E}_{k}$ is given by:
\begin{equation}
\hat{H} = \sum_{k \neq 0} \mathcal{E}_{k} \hat{\beta}^{\dag}_{k} \hat{\beta}_{k},~~~~~~~\mathcal{E}_{k} = \sqrt{\mathcal{A}_{k}^2 - \mathcal{B}_{k}^2} .
\label{H_diag}
\end{equation}

Using the Bogolyubov transformation we can also deduce the quadratic bosonic correlators
$G_k(0) = \langle \hat{b}^{\dag}_k \hat{b}_k \rangle_0$ and $F_k(0) = \langle \hat{b}_{-k} \hat{b}_{k} \rangle_0$, where the notation $\langle ... \rangle_0 = \langle \Psi(0) \vert ... \vert \Psi(0) \rangle$ refers to 
the expectation value with respect to the many-body ground state $\ket{\Psi(0)}$ of the BH chain confined in the SF-mean-field regime.
We find:
\begin{equation}
G_k(0) = v_k^2 = \frac{1}{2}\left(\frac{\mathcal{A}_{k}}{\mathcal{E}_{k}} -1 \right), F_k(0) = u_k v_k = -\frac{\mathcal{B}_{k}}{2\mathcal{E}_{k}}.
\end{equation}

To investigate the dissipative quench dynamics induced by on-site two-body losses of the BH chain initially confined in the SF-mean-field regime, we calculate the equation of motion (EoM) associated to the correlators $G_k(t) = \langle \hat{b}^{\dag}_k \hat{b}_k \rangle_t = \langle \hat{n}_k \rangle_t$ and $F_k(t) = \langle \hat{b}_{-k} \hat{b}_{k} \rangle_t$ using the Lindblad master equation. 
Note that the latter equation will depend on time also via the Hamiltonian in Eq.~\eqref{eq_bhm_quadratic} that becomes time-dependent because of the depleted filling $\bar n(t)$, or equivalently the depleted density $n(t)$, which appears explicitly in the expression of $\mathcal A_k$ and $\mathcal B_k$. For any momentum $k \in \mathcal{B}$ with $\mathcal{B} = [-\pi,\pi]$ being the first Brillouin zone (FBZ), we find:
\begin{widetext}
\begin{subequations}
\label{lindblad_n}
\begin{align}
& \frac{\mathrm{d}}{\mathrm{d}t} G_k(t) = i \left \langle \left[\hat{H}(t), \hat{n}_k \right] \right \rangle_t + \frac{1}{2}\sum_R \left( \left \langle \hat{L}_R^{\dag}\left[\hat{n}_k, \hat{L}_R \right] \right \rangle_t + \mathrm{h.c} \right) ;\\
& \frac{\mathrm{d}}{\mathrm{d}t} F_k(t) = i \left \langle \left[\hat{H}(t), \hat{b}_{-k}\hat{b}_k \right] \right \rangle_t + \sum_R \left \langle \hat{L}_R^{\dag}\hat{b}_{-k}\hat{b}_k \hat{L}_R \right \rangle_t -\frac{1}{2}\left \langle \left\{ \hat{L}_R^{\dag}\hat{L}_R, \hat{b}_{-k}\hat{b}_k \right \} \right \rangle_t,
\end{align}
\end{subequations}
\end{widetext}

\noindent
where $\hat{H}(t)$ depends on $n(t)$ which reads as: 
\begin{equation}\label{eqn:density}
n(t) = \frac{1}{L} \left[G_0(t) + \sum_{k \neq 0} G_k(t)\right]. 
\end{equation}
\noindent
To calculate the EoMs, since the SF-mean-field regime is considered, we rely on a decoupling of the condensate mode $k = 0$ from the other modes together with a mean-field approximation where only the terms depending on correlators involving four or two bosonic operators acting on the mode $k = 0$ are conserved. As a consequence, we have to compute theoretically the EoM associated to $G_k(t)$ and $F_k(t)$ both for $k = 0$ and $k \neq 0$. For the EoM of $G_0(t)$ and $F_0(t)$, the approximations $\langle \hat{n}_0^2\rangle_t = \langle \hat{n}_0 \rangle_t^2$ and $\langle \hat{b}_0 \hat{b}_0 \hat{b}_0^{\dag} \hat{b}_0 \rangle_t = \langle \hat{b}_0 \hat{b}_0 \rangle_t \langle \hat{b}^{\dag}_0 \hat{b}_0 \rangle_t$ have been used. 
Note that these approximations require both a small dimensionless interaction parameter $U/J$, which is satisfied in the SF-mean-field regime under study, as well as small observation times. Finally, after some algebra, we end up with the following set of EoMs:

\begin{widetext}
\begin{subequations}
\label{EoMs}
\begin{align}
 \frac{\mathrm{d}}{\mathrm{d}t} G_0(t) =& -\frac{\gamma}{L}\sum_{q \neq 0}(F_0(t)F_q(t)^* + \mathrm{h.c}) - \frac{4\gamma}{L}G_0(t)\sum_{q \neq 0}G_q(t) - \frac{2\gamma}{L}G_0(t)\left(G_0(t)-1\right) ; \\
 \frac{\mathrm{d}}{\mathrm{d}t} F_0(t) =&  -\frac{\gamma}{L}\left(2G_0(t)-3\right)F_0(t) - \frac{4\gamma}{L}F_0(t)\sum_{q\neq 0}G_q(t) - \frac{\gamma}{L}(2G_0(t)+1)\sum_{q\neq 0} F_q(t) ; \\
 \frac{\mathrm{d}}{\mathrm{d}t} G_k(t) =&  -2\mathcal{B}_{k}(t)\operatorname{Im}(F_k(t)) - \frac{\gamma}{L}\left( F_0(t)F_k(t)^* + \mathrm{h.c} \right) -\frac{4\gamma}{L}G_0(t)G_k(t),~~~~ \forall k \neq 0 ; \\
 \frac{\mathrm{d}}{\mathrm{d}t} F_k(t) =& -\left[2i\mathcal{A}_{k}(t)+\frac{4\gamma}{L}G_0(t)\right]F_k(t) - \left[i\mathcal{B}_{k}(t)+ \frac{\gamma}{L}F_0(t)\right](2G_k(t)+1),~~~~ \forall k \neq 0;
\end{align}
\end{subequations}
\end{widetext}

\noindent
where the initial conditions are given by: 
\begin{subequations}
\label{Eq:Init:0}
\begin{align}
& G_0(0) = N_0 = N - \sum_{k \neq 0} G_k(0); F_0(0) = \Theta_{\rm H}(U) N_0; \label{eq_init_1} \\
& G_k(0) = \frac{1}{2} \left(\frac{\mathcal{A}_{k}}{\mathcal{E}_{k}} - 1 \right); ~~~~~~~~~~~ F_k(0) = -\frac{\mathcal{B}_{k}}{2\mathcal{E}_{k}}. \label{eq_init_2}
\end{align}
\end{subequations}

\noindent
$\Theta_{\rm H}(U)$ denotes the Heaviside function defined as $\Theta(U) = 1$ if $U > 0$ and $\Theta_{\rm H}(U) = 0$ if $U = 0$.
The properties of the EoMs in the special cases $\gamma = 0$, i.e. no sudden global quench, and $U=0$, namely the non-interacting limit, are discussed in Appendix \ref{appendix_prop}. 
In Appendix~\ref{2d_eoms}, the set of EoMs valid for a 2D square lattice is discussed. 

\subsection{EoM approach for long-range two-body losses}
We move on to the case of long-range two-body losses with a power-law decaying dissipation strength $\gamma_{|R-R'|}$. 
Similarly as for on-site two-body losses, we calculate the EoMs associated to the relevant quadratic correlators $G_k(t) = \langle \hat{b}^{\dag}_k \hat{b}_k \rangle_t = \langle \hat{n}_k \rangle_t$ and $F_k(t) = \langle \hat{b}_{-k} \hat{b}_{k} \rangle_t$ using the Lindblad master equation at Eq.~\eqref{lind_mast_eq}. We then find the set of differential equations at Eq.~\eqref{lindblad_n} where the sum over the lattice site index $R$ and the Lindblad jump operator $\hat{L}_R$ are replaced by a double sum over $R$ and $R'$ and by $\hat{L}_{R,R'}$ respectively. Considering the same theoretical approach and approximations used previously in the case of on-site two-body loss processes, 
we obtain the following set of EoMs:
\begin{widetext}
\begin{subequations}
\label{EoMs_nl}
\begin{align}
\frac{\mathrm{d}}{\mathrm{d}t} G_0(t) =& -\sum_{q \neq 0} \left[\left(\mathcal{G}_q+\mathcal{H}_q\right)F_0(t)F_q(t)^* + \mathrm{h.c} \right]
-\sum_{q \neq 0} \mathcal{F}_qG_0(t)G_q(t) - 2\mathcal{G}_{0}G_0(t)(G_0(t)-1); \\
\frac{\mathrm{d}}{\mathrm{d}t} F_0(t) =& - \mathcal{G}_{0}(2G_0(t)-3)F_0(t) -\sum_{q \neq 0}\mathcal{F}_qF_0(t)G_q(t) - \sum_{q \neq 0}
\left(\mathcal{G}_q+\mathcal{H}_q\right)(2G_0(t)+1)F_q(t); \\
\frac{\mathrm{d}}{\mathrm{d}t} G_k(t) =&-2\mathcal{B}_{k}(t)\operatorname{Im}(F_k(t)) - \mathcal{G}_k(F_0(t)F_k(t)^* + \mathrm{h.c}) - \mathcal{F}_kG_0(t)G_k(t),~~~~ \forall k \neq 0; \\
\frac{\mathrm{d}}{\mathrm{d}t} F_k(t) = &-\left[2i\mathcal{A}_{k}(t) + \mathcal{F}_kG_0(t)\right]F_k(t) -\left[i\mathcal{B}_{k}(t) + \mathcal{G}_k F_0(t) \right](2G_k(t)+1),~~~~ \forall k \neq 0,
\end{align}
\end{subequations}
\end{widetext}

\noindent
where the momentum-dependent functions $\mathcal{F}_q$, $\mathcal{G}_q$ and $\mathcal{H}_q$ are defined as follows:
\begin{subequations}
\begin{align}
 \mathcal{F}_q = & ~ 2[\mathcal{G}_{0}+\mathcal{G}_q];\\
 \mathcal{G}_q = & ~ \frac{1}{L^2}\sum_{R,R'}\gamma_{|R-R'|}\cos(q(R-R')),\label{gamma_q} \\
 \mathcal{H}_q = & ~ \frac{i}{L^2}\sum_{R,R'}\gamma_{|R-R'|}\sin(q(R-R')).
\end{align}
\end{subequations}

Note that the same initial conditions as in the case of on-site two-body losses are considered, see Eqs.~\eqref{Eq:Init:0}. In Appendix~\ref{appendix_non_local}, we discuss the main properties of the set of EoMs valid for long-range two-body losses.

\subsection{Tensor networks numerical calculations}
All the numerical results presented below are obtained using the quantum jump method based on quantum trajectories, see Ref.~\cite{daley2014} and 
Appendix~\ref{appendix_quantum_jump}, in the framework of tensor networks~\cite{schollwock2005,schollwock2011} where the Julia version of the ITensor
package has been used~\cite{itensor,itensor2}. For all the considered cases, a precise analysis of the cutoffs has been systematically performed to ensure the convergence of the numerical results. These cutoffs include the dimension of the local Hilbert space $\mathbb{H}_R$, where we considered $\mathrm{dim}(\mathbb{H}_R) = N(0)+1$ with $N(0)$ the total number of bosonic particles initially present on the lattice. It also comprises the number of quantum trajectories $N_{\mathrm{traj}}$ with $N_{\mathrm{traj}} \in [10^3, 3\times 10^3]$, the bond dimension $\chi$ of the truncated time-evolved matrix product state $\ket{\Psi(t)}$ with $\mathrm{max}(\chi) = 5\times 10^2$ and the time-step $\delta t$ where $\delta t = \{2.5\times 10^{-2}, 5\times 10^{-2}\}$. 
The latter are critical when the initial many-body quantum state $\ket{\Psi(0)}$ corresponds to the ground state of the BH model confined in the 
SF-mean-field regime. Indeed, the numerical requirements are the most binding in this specific regime when comparing to the strongly-interacting regime of the SF phase or the MI phase.
This can be explained by the strong coherence, thus implying strong entanglement entropy, due to the large density fluctuations. Finally, we stress that 
the length and time scales considered for the numerical simulations are comparable to those reached experimentally in ultra-cold-atom experiments.  

\section{Numerical Results for Local Losses}
\label{sec:BH_chain}
In this section, we investigate the dissipative quench dynamics induced by on-site two-body loss processes of the BH chain. The equilibrium phase diagram of the model contains a superfluid phase  (SF) and a Mott insulating (MI) phase separated by a quantum phase transition at $(U/J)_{\mathrm{c}}^{\bar{n} = 1} \simeq 3.3$. We focus here on the weakly-interacting SF regime.

\subsection{EoM benchmark: weakly-interacting case}
\begin{figure}[t]
\includegraphics[scale = 0.35]{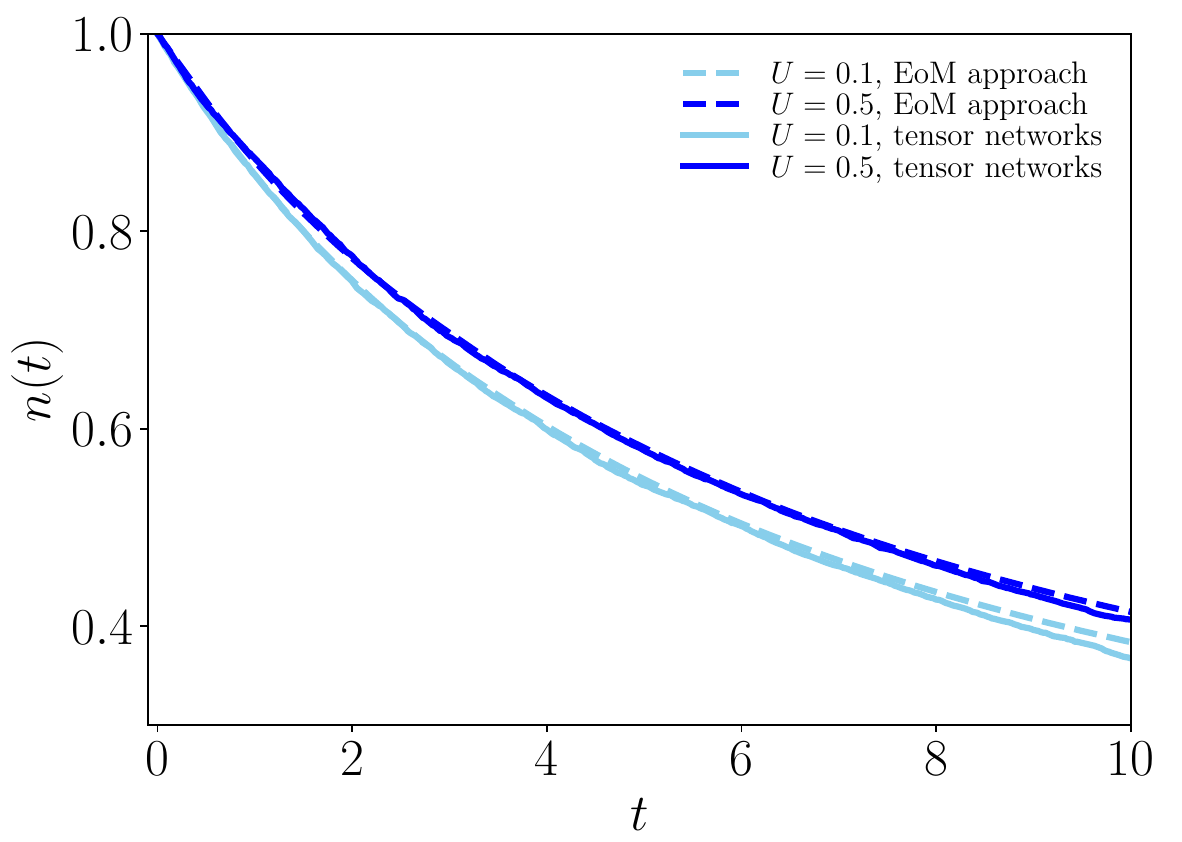}
\caption{Bosonic density $n(t)$ as a function of time $t$ for a sudden global quench on the dissipation strength 
from $\gamma = 0$ to $\gamma = 0.1$ of the BH chain initially confined in the SF-mean-field regime and submitted to on-site two-body losses. The solid lines represent numerical results obtained from the quantum jump method using tensor networks whereas the dotted lines correspond to theoretical predictions from the EoM approach given at Eq.~\eqref{EoMs}. The parameters are: $N(0) = L = 12$, $J=1$.}
\label{finite_interaction}
\end{figure}

In the SF-mean-field regime, we expect our EoM approach based on Bogolyubov theory to work. We start therefore by benchmarking the dissipative dynamics obtained by integrating Eq.~\eqref{EoMs} against
tensor-network-based numerical calculations relying on the quantum jump approach for different values of $U$ and small system sizes, i.e.~$L \in \{12,18\}$. On Fig.~\ref{finite_interaction} we plot the dynamics of the particle density $n(t)$ for two distinct interaction strengths, namely $U=0.1 J$ and $U=0.5 J$. We see that both approaches agree qualitatively and quantitatively and display a decay of the particle density, which is slightly slower for larger values of interaction. This effect can be explained by the decreasing number of multi-occupied lattice sites initially and during the quench dynamics. The latter benchmark permits to show the robustness of our theoretical approach since the mean-field condition, $\bar{n} \gg U/J$, is not verified satisfactorily. 

\begin{figure}[t]
\includegraphics[scale = 0.35]{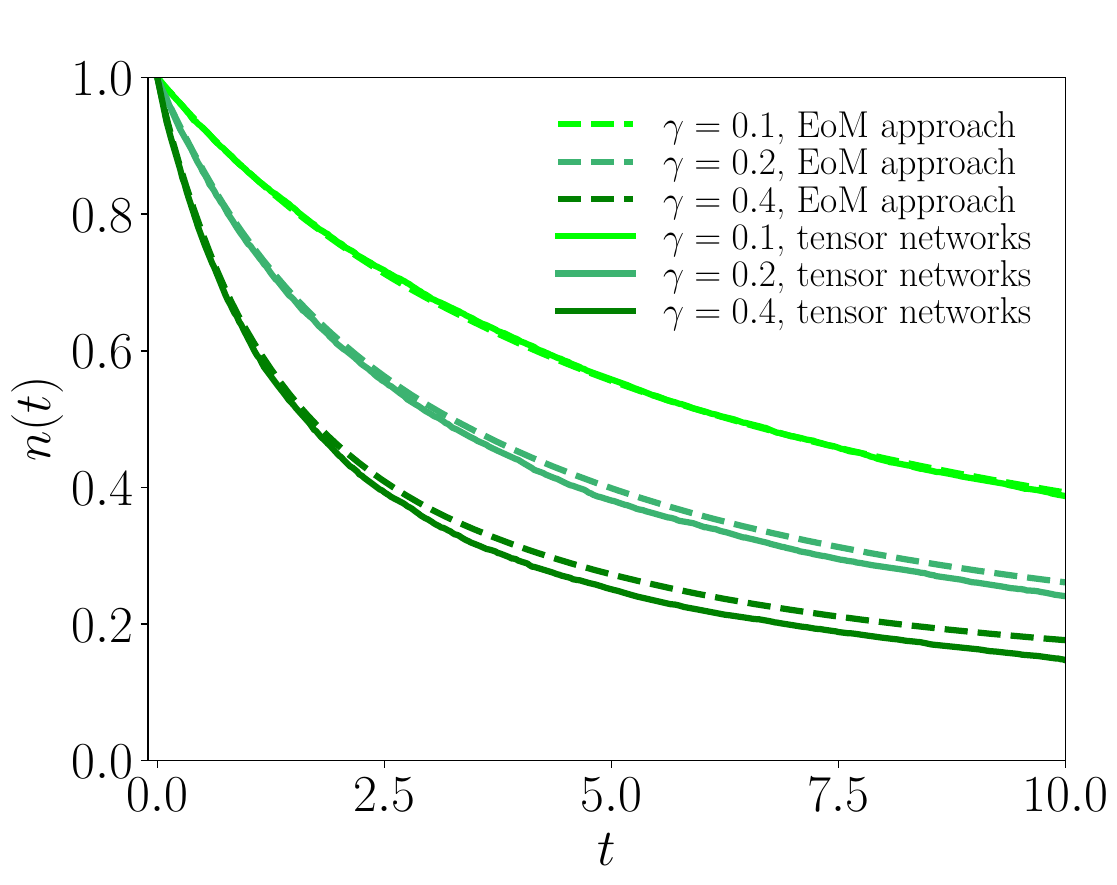}
\caption{Bosonic density $n(t)$ as a function of time $t$ for a sudden global quench on the dissipation strength 
from $\gamma = 0$ to $\gamma > 0$ for a fixed two-body repulsive interaction strength $U$ of the BH chain initially confined in the SF-mean-field regime and submitted to on-site two-body losses. The solid lines represent numerical results obtained from the quantum jump method using tensor networks whereas the dashed lines correspond to theoretical predictions from the EoM approach given at Eq.~\eqref{EoMs}. The parameters are: $N(0) = L = 12$, $J = 1$, $U = 0.2$.}
\label{gamma_study}
\end{figure}

We turn to a similar study where the interaction strength $U$ is fixed while the dissipation strength $\gamma$ varies. From Fig.~\ref{gamma_study}, we clearly show a very good agreement between our theoretical EoM approach and the numerical simulations using the quantum trajectory method. We can also notice a stronger decay of $n(t)$ when increasing $\gamma$, which is expected. 

Both studies, where the $U$-dependence as well as the $\gamma$-dependence of $n(t)$ have been investigated, permit to validate our theory, at least in the SF-mean-field regime. In Appendix~\ref{app:additional}, we provide an additional benchmark between theoretical and numerical results where the previous study regarding the $\gamma$-dependence is performed for a larger chain length $L$ at unit-filling. Note that
the investigation concerning $U$-dependence is not shown but has been done and a very good agreement has been found. 

\subsection{Long-time decay of the particle density}
\label{scaling_law_local}
We now employ the theory based on the EoMs in Eq.~\eqref{EoMs} to investigate the time-dependent density profile $n(t)$ for a sudden global quench
on the dissipation strength $\gamma$ for the BH chain initially confined in the weakly-interacting SF-mean-field regime; we consider intermediate
observation times $T$, i.e. $\mathrm{max}(T) = 50$ (in units of $J^{-1}$) and a large system size $L$, i.e. $L = 100$ (in units of $a$), in order to get a high initial number of bosons since a unit-filling of the lattice chain is considered. For various values of the dissipation and interaction strengths denoted by $\gamma$ and $U$ respectively, we characterize the decay in time of the density as a function of $U$ at fixed $\gamma$ and conversely. 

\begin{figure*}[t]
\includegraphics[scale = 0.46]{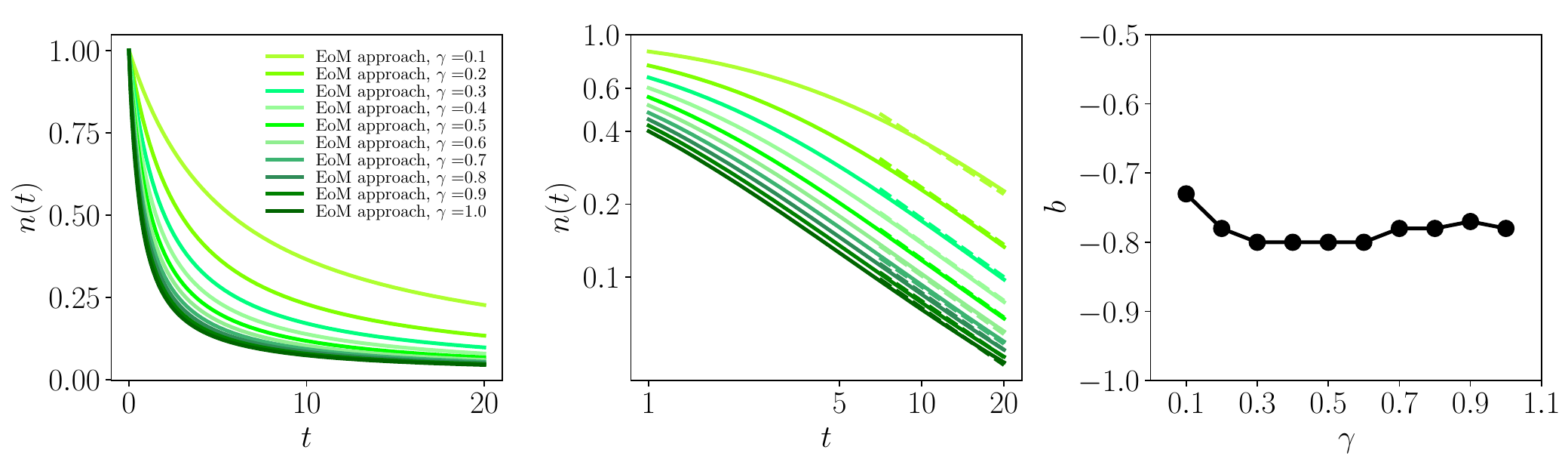}
\caption{Bosonic density $n(t)$ as a function of time $t$ for the BH chain confined initially in the SF-mean-field regime for a sudden global quench on the dissipation strength $\gamma$ from $\gamma = 0$ to $\gamma > 0$. The latter is investigated for various $\gamma$ while fixing the interaction strength $U$. (Left) time-dependent density profile in linear-linear scale (center) in log-log scale. The solid lines represent theoretical results obtained from the EoM approach whereas the dashed lines are the corresponding fits having an ansatz of the form $n_{\mathrm{fit}}(t) = at^b$. (Right) power-law exponent $b$ as a function of $\gamma$. The parameters are: $N(0) = L = 100$, $J = 1$, $U = 0.1$.} 
\label{fig_fits_U_small_gamma_varies}
\end{figure*}

First, we consider $U = 0.1 J$ while $\gamma$ varies as shown in Fig.~\ref{fig_fits_U_small_gamma_varies} and plot the dynamics of the particle density. The short-time transient depends strongly on the value of $\gamma$, however on longer time scales the density displays a power-law decay which appears to be independent on the dissipation strength. To verify this point we fit the decay in time using an algebraic function of the form $n_{\mathrm{fit}}(t) = at^b$. 
According to the values of the pre-factor $a$ and the power-law exponent $b$ for the different fits, see Fig.~\ref{fig_fits_U_small_gamma_varies}, the density profiles are shifted while the algebraic decay in time remains unchanged and is characterized by a power-law exponent $b \simeq -0.8$. A similar study for a larger interaction strength, i.e. $U = 0.3 J$, is reported in Appendix~\ref{app:additional} with similar qualitative results, namely an exponent weakly dependent on dissipation, except for
small $\gamma$. The latter point, i.e. the $U$-dependence of $n(t)$ at small $\gamma$, is discussed theoretically later on. 

\begin{figure*}[t]
\includegraphics[scale = 0.44]{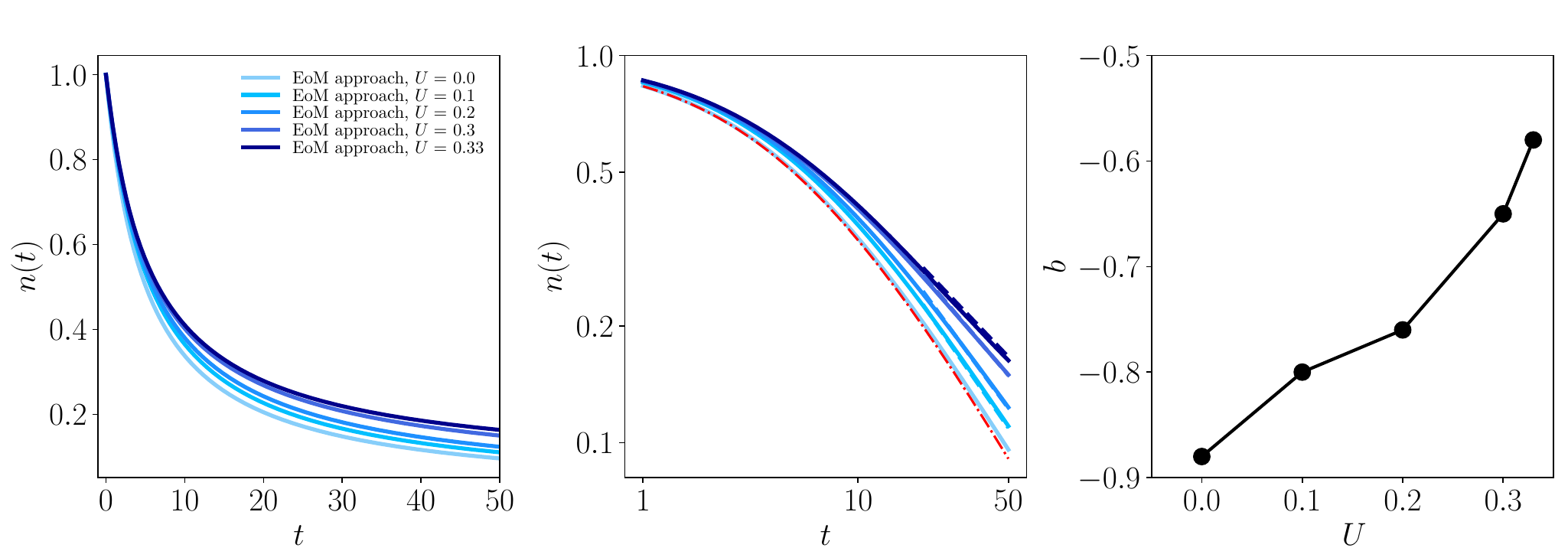}
\caption{Bosonic density $n(t)$ as a function of time $t$ for the BH chain initially confined in the SF-mean-field regime for a sudden global
quench on the dissipation strength $\gamma$ from $\gamma = 0$ to $\gamma > 0$. The latter is investigated for various interaction strengths $U$ while fixing $\gamma$. (Left) time-dependent density profile in linear-linear scale (center) in log-log scale. The solid lines represent
theoretical results obtained from the EoM approach whereas the dashed lines are the corresponding fits having an ansatz 
of the form $n_{\mathrm{fit}}(t) = at^b$ except for the dashed-dotted red line representing the analytical expression of $n(t)$ for the non-interacting case in the thermodynamic limit, see Eq.~\eqref{eq_density_U_0}. (Right) power-law exponent $b$ as a function of $U$. The parameters are: $N(0) = L = 100$, $J = 1$ and $\gamma = 0.1$.}
\label{fig_fits_U_varies_gamma_small}
\end{figure*}

A power-law decay of the bosonic density is expected in the non-interacting case, where according to mean-field theory one obtains in the thermodynamic limit the form
$n(t) = \bar n(0) / (1+ 2 \gamma \bar n(0) t)$ (See Appendix~\ref{appendix_prop}). Here instead, quite remarkably, we obtain a power-law decay with a different exponent. As we are going to show next, this difference is due to interactions. To demonstrate this point, we fix $\gamma$ and change the value of $U$ to study the decay of particle density as shown on Fig.~\ref{fig_fits_U_varies_gamma_small}. We see that the interaction affects less the short-time transient now, while they seem to control the long-time decay which is consistent with a power-law with an interaction-dependent exponent, see right panel of Fig.~\ref{fig_fits_U_varies_gamma_small}. In particular, the density dynamics slow down upon increasing the interaction. The previous physical property requires $\gamma \simeq U \ll J$; this in order to maintain a balance between $U$-driven and $\gamma$-driven dynamics. Indeed, as previously shown when large $\gamma$ were considered, the dissipation completely dominates the dynamics and thus the interaction dependence can be disregarded. We will come back to this point in the next Section.

The numerics in the non-interacting case is also compared to the infinite-size limit result $n(t) = \bar n(0) / (1+ 2 \gamma \bar n(0) t)$, which is plotted with a dashed-dotted red line in Fig.~\ref{fig_fits_U_varies_gamma_small}. A very good agreement is visible; however, the extracted exponent $b(U)$ approaches the value $-0.9$  for $U\rightarrow0$, instead of the value $-1$ as expected. This difference is due to a finite-size effect. Indeed, for finite $L$, the particle density decay reads as in Eq.~\eqref{N_U_0}, which is reported here for better readability:
\begin{equation}
n(t) = \frac{1/L}{1-e^{-(at+c)}},~a = \frac{2\gamma}{L},~c = \mathrm{ln}\left(\frac{N(0)}{N(0)-1}\right) .
\end{equation}
In other words, for finite $L$, the decay of the density at long times is in fact exponential and the power-law regime is only present at intermediate times which makes the fitting procedure less reliable. 

To summarize, we have found  that the long-time dynamics of the particle density in the weakly interacting case still displays a power-law decay, but with an exponent that
at weak dissipation depends on interaction. More precisely when $U$ increases then $|b|$ decreases, i.e. interactions slow down the depletion of the gas due to two-body losses. While this effect at short time could be attributed to the reduced double occupancy in the initial state, controlling the rate of decay of the total particle number, here we highlight that this interaction-dependence affects the long-time power-law decay of the density. In the next section we will try to elaborate further on this result.
Finally,  It is natural to wonder if this specific behavior of $n(t)$ is still valid for larger $\gamma$. This is not the case as shown in Appendix~\ref{app:additional} where the quench dynamics is determined by the on-site dissipation strength $\gamma$, without any significant dependence on the interaction strength $U$.

\subsection{Pseudo-closed equation for $n(t)$ and correlation functions} 
The power-law decay result obtained in the non-interacting case can be derived from a simple and transparent self-consistent equation for the particle density, given by $\mathrm{d}n(t)/\mathrm{d}t = -\gamma n(t)^2$. 
It is therefore tempting to imagine whether a similarly simple equation could be derived for the particle density in the presence of a finite interaction strength $U$, while using the assumptions behind Bogolyubov theory.

We present here the derivation of a pseudo-closed equation (PCE) for the quench dynamical behavior of the density $n(t)$ in the weakly-interacting SF-mean-field regime. 
The basic idea  consists in taking advantage of the EoMs in Eq.~\eqref{EoMs} and substitute them in the expression from the time-derivative of the particle density, from Eq.~(\ref{eqn:density}). We first consider the thermodynamic limit  and notice that $\operatorname{Re}(F_0(t)) \gg |\operatorname{Im}(F_0(t))|$ and 
$|\sum_{q\neq0}\operatorname{Re}(F_q(t))| \gg |\sum_{q\neq0} \operatorname{Im}(F_q(t))|$. Then, we conclude that the EoM associated to $\operatorname{Re}(F_0(t))$ and $G_0(t)$ are the same and since the initial conditions are the same,
i.e. $\operatorname{Re}(F_0(0)) = G_0(0)$, we
get $\operatorname{Re}(F_0(t)) = G_0(t)$. Once we replace $\operatorname{Re}(F_0(t))$ by $G_0(t)$ in the EoM for $N(t)$, we use the mean-field approximation $N(t)\gg \sum_{k\neq0}G_k(t)$ to obtain a PCE for the depleted density $n(t)$, which reads:
\begin{equation}
 \frac{\mathrm{d}}{\mathrm{d}t} n(t) = -2\gamma n(t)^2  -4\gamma n(t)\frac{1}{L}\sum_{q\neq0} \left[G_q(t) + \operatorname{Re}(F_q(t))\right].
\label{eq_pce}
\end{equation}
In Appendix~\ref{appendix_prop} we present a second theoretical approach to deduce the PCE associated to $n(t)$.

In order to certify Eq.~\eqref{eq_pce}, we have verified that it reproduces the density dynamics computed from the EoMs at Eq.~\eqref{EoMs}. 
Note that the approximations considered previously imply $U/J \rightarrow 0$, i.e.~to be well confined in the SF-mean-field regime, and to consider relatively small observation times $T$. 
These conditions are already satisfied since the starting point of our calculations is the set of EoMs at Eq.~\eqref{EoMs} based on similar requirements. 
To further validate Eq.~\eqref{eq_pce}, it is also interesting to discuss some limiting behaviors.
In the non-interacting limit, namely $U = 0$, we have $G_q(t) = F_q(t) = 0$ for $\forall q \neq 0$ and $\forall t$. Hence, we get $\mathrm{d}n(t)/\mathrm{d}t = -2\gamma n(t)^2$ and we recover the result we discussed before.
In the non-dissipative limit, namely $\gamma = 0$, we find $\mathrm{d}n(t)/\mathrm{d}t = 0$ as expected. 

Let us now discuss the physical meaning of Eq.~\eqref{eq_pce}. The correction term corresponding to the second term on the right-hand side of the equation is due to the finite interaction strength $U$.  Indeed within Bogolyubov theory interaction $U$ has two main effects: (i) it scatters particles out of the condensates leading to finite occupation of modes with $k\neq0$, i.e. $G_k(t)\neq0$ and (ii) it squeezes the modes outside of the condensate (see Eq. ~\ref{eq_bhm_quadratic}) leading to a finite value for the off-diagonal correlation $F_{q}(t)$. We now note from the numerical solution of the EoM that these terms satisfy $\sum_{q\neq0} G_q(t) > 0$, $\sum_{q\neq0} \operatorname{Re}(F_q(t)) < 0$ such that $|\sum_{q\neq0} \operatorname{Re}(F_q(t))| > \sum_{q\neq0} G_q(t)$. This property becomes more and more valid when increasing $U$. From Eq.~\eqref{eq_pce}, the latter implies that the effective decay rate of $n(t)$ decreases when increasing $U$, and thus at a fixed time $t$ we should expect a larger density while $U$ is increased. The previous statement is verified in Fig.~\ref{finite_interaction}. 
Overall, this suggests a simple physical interpretation of Eq.~\eqref{eq_pce}. In presence of interactions, when not all particles are in the condensate, the rate of depletion of the system due to two-body losses takes contributions both from dissipative processes within the condensate or within the modes at finite $k$ outside of the condensate or even between one particle in the condensate and one outside of it.

It is interesting to compare the PCE with existing results in the literature. In particular, Refs.~\cite{liu2022,liu2024} studied the related problem of a lossy bosonic condensate and obtained a closed self-consistent equation
for the particle density. The main difference with respect to our approach is that here  we used a time-dependent Bogolyubov theory as well as the mean-field approximation in order to deduce the PCE associated to $n(t)$. 
On the other hand,  Refs~\cite{liu2022,liu2024} approximate $N(t)$ with the occupation number of the condensate mode $k = 0$, i.e. $N(t) = G_0(t)$, and disregard the coupling between  the correlators in the mode $k = 0$ and the modes $k \neq 0$. In addition, the dynamics of the quadratic bosonic correlators $G_k(t)$, $\operatorname{Re}(F_k(t))$ and $\operatorname{Im}(F_k(t))$ are also different between the two approaches. Here we consider the full self-consistent dynamics arising from a time-dependent density, while in Refs.~\cite{liu2022,liu2024}  the density is assumed to be constant when solving for the dynamics of the bosonic correlators. These assumptions are very binding and require to consider very short observation times $T$ to be valid, while our approach compares well with quasi-exact tensor-network-based simulations up to long time scales.

\section{Numerical Results for Long-Range Losses}
\label{sec:non_local}
We move on to the case of non-local loss processes and more precisely to long-range two-body losses characterized by a dissipation strength displaying an algebraic decay spatially. \\

\begin{figure}[t]
\includegraphics[scale = 0.41]{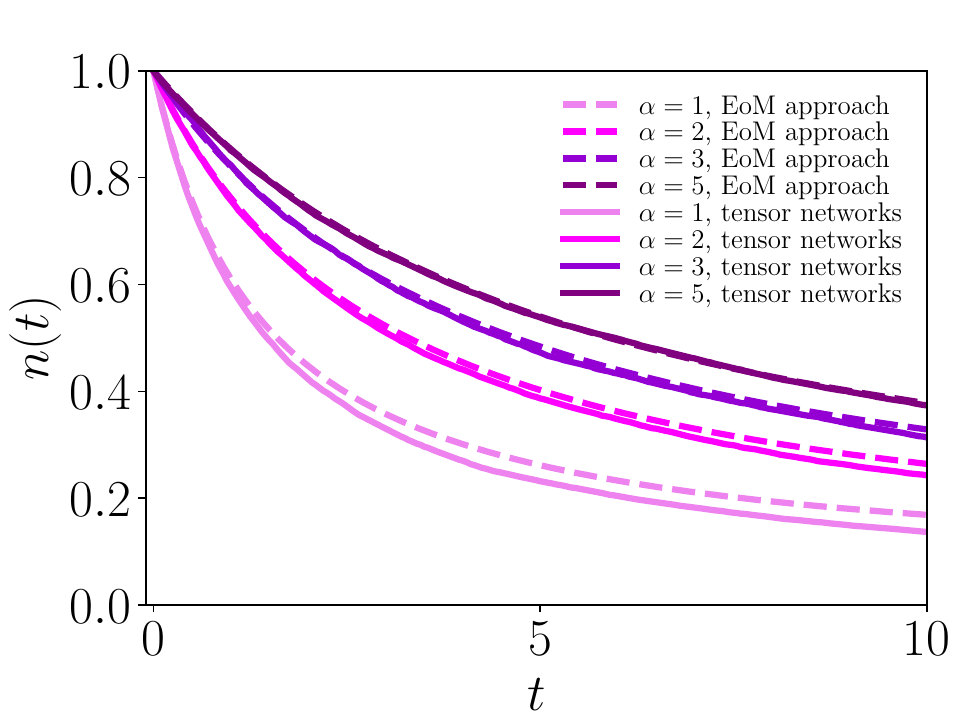}
\caption{Bosonic density $n(t)$ as a function of time $t$ of the BH chain initially confined in the SF-mean-field regime for a sudden global quench on the dissipation strength from $\Gamma = 0$ to $\Gamma > 0$ where various values of the power-law exponent
$\alpha$ are considered. The solid lines represent numerical results obtained from the quantum jump method using tensor networks whereas the dashed lines correspond to theoretical predictions obtained from the EoMs at Eq.~\eqref{EoMs_nl}. The parameters are: $N(0) = L = 12$, $J=1$, $U = 0.2$,
$\Gamma = 0.1$.}
\label{fig_alpha}
\end{figure}

In what follows, we investigate $n(t)$ for a sudden global quench on the dissipation strength from $\Gamma = 0$ to $\Gamma > 0$ for the BH chain initially confined 
in the SF-mean-field regime. We compare our theoretical predictions obtained from the EoMs at Eq.~\eqref{EoMs_nl} with tensor networks numerical calculations based on the quantum trajectory method, see Appendix~\ref{appendix_quantum_jump}. The latter investigation is performed for various power-law exponents $\alpha$, for various dissipation strengths $\Gamma$ for a fixed value of $\alpha$.

On Fig.~\ref{fig_alpha}, we benchmark our theoretical results with numerical calculations where a very good agreement is found. The error increases when $n(t)$ reaches small values which was expected to occur since the EoMs are based on the mean-field approximation implying $N(t) \gg \sum_{k \neq 0} G_k(t)$ which is fulfilled for relatively large values of $N(t)$. The latter error becomes larger when decreasing $\alpha$ implying a stronger long-range dissipation which permits to reach smaller values of $n(t)$ for a same time $t$. 

\begin{figure}[t]
\includegraphics[scale = 0.41]{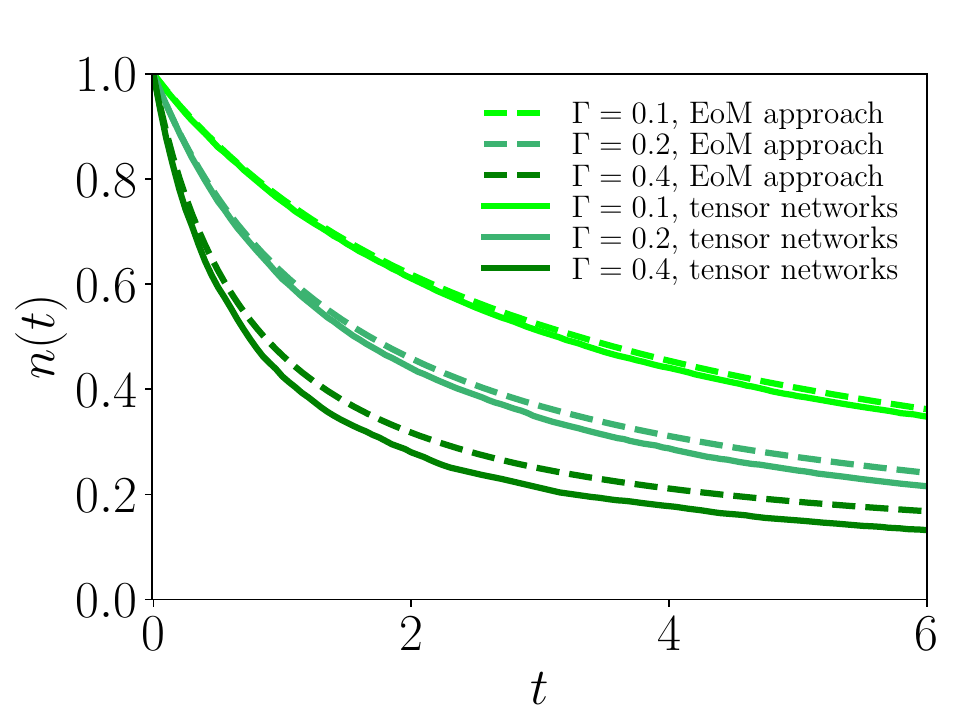}
\caption{Bosonic density $n(t)$ as a function of time $t$ of the BH chain initially confined in the SF-mean-field regime for a sudden global quench on the dissipation strength from $\Gamma = 0$ to $\Gamma > 0$ where various values of the post-quench dissipation strength $\Gamma$ are considered for a fixed value of $\alpha$. The solid lines represent numerical results obtained from the quantum jump method using tensor networks whereas the dashed lines correspond to theoretical predictions obtained from the EoMs at Eq.~\eqref{EoMs_nl}. The parameters are: $N(0) = L = 12$, $J = 1$, $U = 0.2$,
$\alpha = 2$.}
\label{fig_gamma_maj}
\end{figure}

On Fig.~\ref{fig_gamma_maj}, we investigate the $\Gamma$-dependence of $n(t)$ while fixing $\alpha$ and where we benchmark our EoM approach with tensor-network-based numerical calculations. 

Finally, we studied the $U$-dependence of $n(t)$ for small values where our theory reproduces accurately the numerical results. However, this benchmark between the two approaches is limited to small interaction strengths $U$ such that the BH chain is initially well confined in the SF-mean-field regime. Indeed, for larger $U$, our EoM approach is expected to fail.
This can be explained by the fact that the BH chain is no longer confined in the SF-mean-field regime but rather in the strongly-correlated regime of the SF phase where the approximations considered during the derivation of the EoMs are not satisfied anymore. Indeed, at relatively large $U$, the decoupling between the condensate mode $k=0$ and the finite modes $k \neq 0$ as well as the product state approximation in momentum space, i.e. $\langle \hat{n}_0^2 \rangle_t = \langle \hat{n}_0 \rangle_t^2$, are no longer valid. \\

We provide here further details on the density profile for long-range two-body losses while considering a large system size, i.e. $L = 100$, to investigate its intermediate-time dynamics. For large power-law exponents $\alpha$, i.e. $\alpha \gg 1$, the behaviour in time of $n(t)$ will be identical to the one for on-site two-body losses discussed in Sec.~\ref{scaling_law_local}. In what follows, we discuss the case of strong long-range loss processes, i.e. $ 0 < \alpha \lesssim 5$. According to the set of EoMs at Eq.~\eqref{EoMs_nl}, Eq.~\eqref{density_nl_non_int} and \eqref{density_nl_pce}, the dissipation strength for long-range algebraic loss processes is characterized by the effective dissipation rate  $\tilde{\gamma}(0) = L\mathcal{G}_0$, see also Eq.~\eqref{gamma_q}, and not $\Gamma$ corresponding to the on-site dissipation strength in the framework of long-range two-body losses. Consequently, for
$\tilde{\gamma}(0) \simeq \gamma$, we expect for $n(t)$ to behave similarly to the case of on-site two-body losses at dissipation rate $\gamma$. 

\begin{figure}[t]
\includegraphics[scale = 0.48]{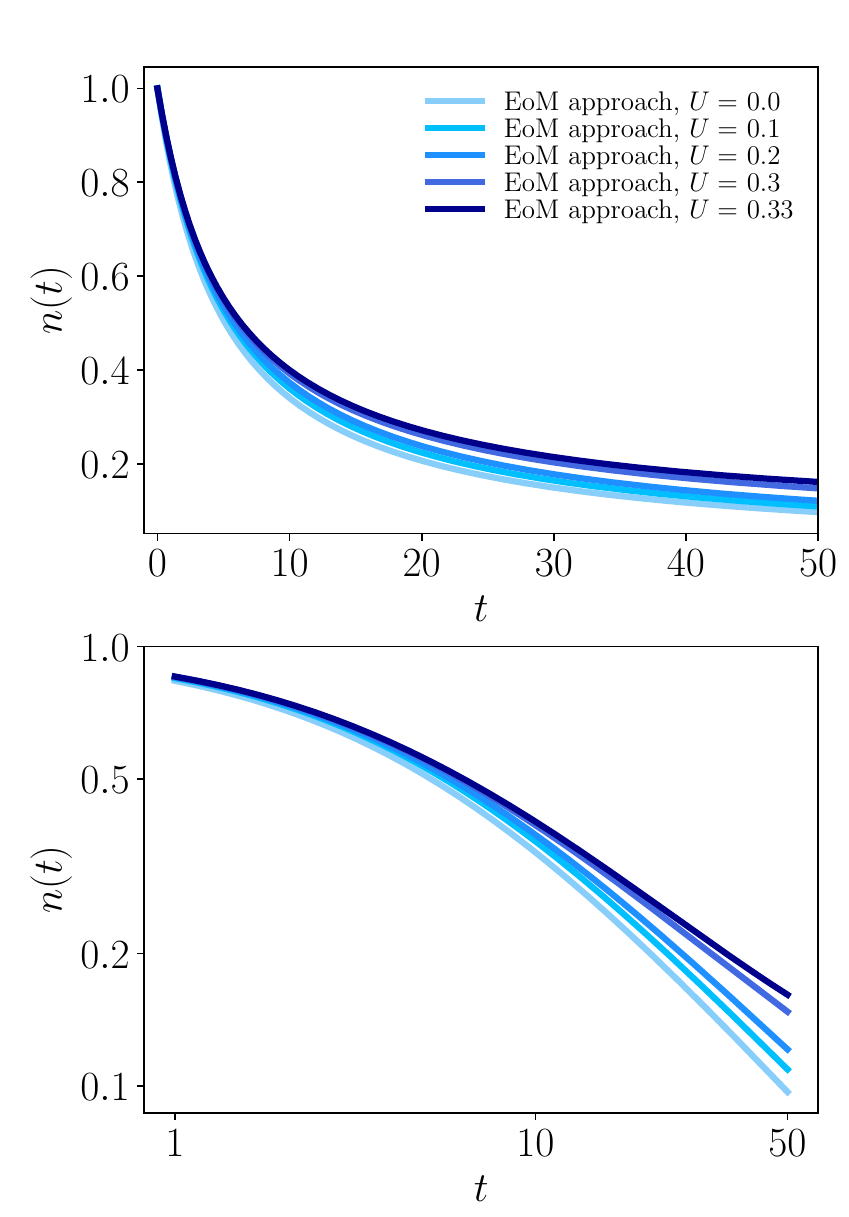}
\caption{Time-dependent density $n(t)$ for the BH chain initially confined in the SF-mean-field regime with a sudden global quench on the dissipation strength from $\Gamma = 0$ to $\Gamma > 0$ where various values of the two-body repulsive interaction strength $U$ are considered at a fixed and small value of $\Gamma$. The solid lines correspond to theoretical predictions obtained from the EoMs at Eq.~\eqref{EoMs_nl}. The parameters are: $N(0) = L = 100$, $J = 1$, $\alpha = 2$ and $\Gamma = 4.5\times 10^{-2}$ leading to $\tilde{\gamma}(0) \simeq 0.1$.}
\label{fig_scaling_law_nl}
\end{figure}

To verify the latter statement, we focus on the interesting case where both $U$ and $\Gamma$ are small, i.e.\ $U,\Gamma \ll J$. 
On Fig.~\ref{fig_scaling_law_nl}, the on-site dissipation rate $\Gamma$ is chosen such that $\tilde{\gamma}(0) \simeq 0.1$. As expected, we recover very similar theoretical predictions as those presented on Fig.~\ref{fig_fits_U_varies_gamma_small} valid for on-site two-body losses with a dissipation strength $\gamma = 0.1$ where the algebraic decay in time is characterised by a $U$-dependent power-law exponent. The time behavior of the density $n(t)$ for long-range two-body losses has also been investigated for a large dissipation strength $\Gamma$ and for distinct small interaction strengths $U$. A strong on-site dissipation strength $\Gamma$ implies a large effective dissipation rate $\tilde{\gamma}(0)$ and we expect for
the different densities to display an algebraic decay in time and most importantly to collapse on the same curve leading to a $U$-independent power-law exponent similarly to the case of on-site two-body losses at large $\gamma$ as depicted in Fig.~\ref{fig_fits_U_varies_gamma_large}. This statement has been verified but is not presented here.   

\section{Lossy Dynamics in two-dimensions}
\label{sec:2D}
In this final section we extend our numerical results for the lossy dynamics of the Bose-Hubbard model to two-dimensional lattices. For simplicity we consider only the case of local losses. We emphasize that the dynamics of the 2D Bose-Hubbard model is very challenging to solve and efficient numerical methods are missing. For this reason, we will limit here to the EoM approach that can be naturally extended to the 2D case. \\

\begin{figure}[t]
\includegraphics[scale = 0.49]{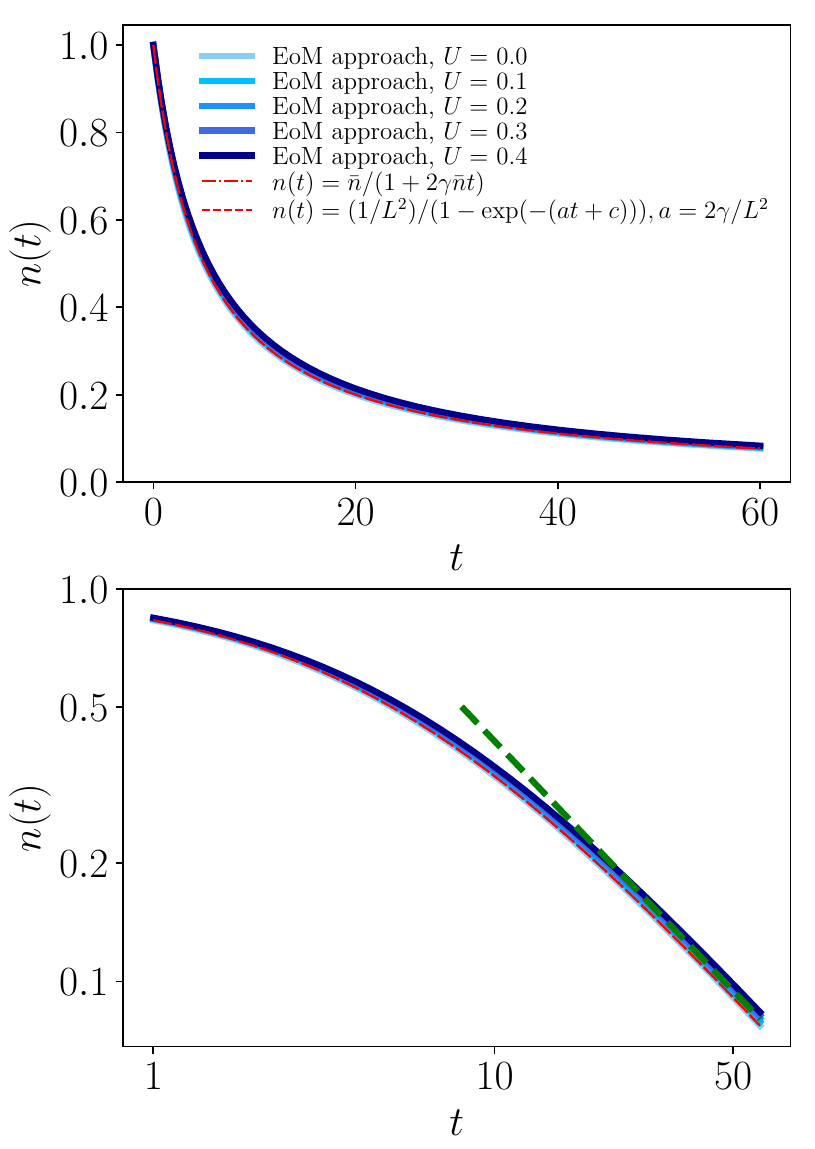}
\caption{Bosonic density $n(t)$ as a function of time $t$ of the 2D BH model on a square lattice and initially confined in the SF-mean-field regime for a sudden global
quench on the dissipation strength from $\gamma = 0$ to $\gamma > 0$. The latter is investigated for various interaction strengths $U$ while fixing $\gamma$.
(Top) time-dependent density profile in linear-linear scale (bottom) in log-log scale. The solid blue lines represent theoretical results obtained from the EoM approach, the dashed green line corresponds to an algebraic fit of the form $n_{\mathrm{fit}}(t) = at^b$ with $a = 3.33$ and $b = -0.91$. The red lines consist in theoretical results for the non-interacting case $U = 0$. The dashed red line corresponds to the analytical expression of the density $n(t)$ deduced from Eq.~\eqref{eq_density_U_0_2D_first} whereas the dashed-dotted red line refers to Eq.~\eqref{eq_density_U_0_2D}. The parameters are: $N(0) = L^2 = 80^2$, $J = 1$, $\gamma = 0.1$.}
\label{fig_fits_U_varies_gamma_small_2D}
\end{figure}

Relying on the EoMs provided at Eq.~\eqref{eq_2D_EoMs}, we investigate here the time-dependent density $n(t)$ for a sudden global quench on the dissipation strength
$\gamma$ for the 2D BH model on a square lattice and initially confined in the SF-mean-field regime. More precisely, we characterize the decay in time of $n(t)$ as a function of $U$ for a fixed and small value of $\gamma$. The EoM approach predicts a $U$-independent density profile in time as shown on Fig.~\ref{fig_fits_U_varies_gamma_small_2D}. Indeed, the decay in time of $n(t)$ at finite $U$ remains unchanged compared to the non-interacting case. This property of $n(t)$ valid for a 2D square lattice drastically differs from the 1D case where the algebraic decay in time of $n(t)$ is characterized by a $U$-dependent power-law exponent, see Fig.~\ref{fig_fits_U_varies_gamma_small}. The physical explanation of the latter property lies in the fact that the critical point of the SF-MI quantum phase transition in 2D is much larger than the one in 1D, i.e. $(U/J)_{\mathrm{c,2D}}^{\bar{n}=1} \gg (U/J)_{\mathrm{c,1D}}^{\bar{n}=1} \simeq 3.3$. Consequently, the range of interaction strengths $U$ considered here is too small compared to $(U/J)_{\mathrm{c,2D}}^{\bar{n}=1}$ to have an effect on the dynamics and thus its dependence can be disregarded. In Appendix~\ref{appendix_dim}, we also unveil by simple theoretical arguments the influence of the lattice dimensionality on the decay in time of $n(t)$. To recover a $U$-dependence for the dynamics, we can drastically increase the latter. However, the 2D BH model will no longer be confined in the SF-mean-field regime since the condition $\bar{n} \gg U/J$ is not satisfied anymore and thus our EoM approach will fail. From a numerical point of view, this study is also complex. Indeed, one possibility to solve the latter problem would be to increase significantly the initial 
filling $\bar{n}$ of the lattice to verify satisfactorily the previous mean-field condition. However, this implies to increase significantly as well the dimension of the local Hilbert space to capture accurately the dissipative dynamics. 

We performed a similar investigation for a stronger quantum quench, i.e.\ when considering a larger dissipation strength $\gamma$. We found that the $U$-independent decay in time of $n(t)$ remains valid even at large $\gamma$. The latter characteristic is reminiscent of the one unveiled for the BH chain in the presence of large on-site two-body losses, see Appendix~\ref{app:additional}.  

\section{Conclusion}
\label{sec:conclusion}

In this work we have discussed the dissipative quench dynamics induced by on-site or long-range two-body losses of the BH model 
on a lattice chain, a 2D square lattice as well as on a hypercubic lattice theoretically and numerically. 

We have first introduced our theory based on an EoM approach and applied the latter to the SF-mean-field regime of the BH chain for 
on-site two-body losses. For various interaction and dissipation strengths, we have shown that this theoretical approach works very well by comparing its predictions against tensor networks numerical results obtained using the quantum jump method. We have also shown that in this regime the intermediate-time dynamics of the density displays an algebraic decay characterized by an interaction-dependent power-law exponent.  

Then, we moved on to a similar investigation for long-range two-body losses. We have certified the validity of our EoM-based theory 
for the BH chain initially confined in the SF-mean-field regime by comparing its predictions by numerical calculations relying on the same method than previously. This has been performed for various values of both the power-law exponent and the dissipation strength. 

Finally, we investigated the dissipative quench dynamics of the 2D BH model for on-site two-body losses in the SF-mean-field regime theoretically, for weak-interactions. We found that the time-dependent behavior of the density is interaction- and dissipation-independent and remains very similar to the non-interacting case, at least for the range of interactions we have explored.  The dependence on the lattice dimensionality of the density profile has been explained using physical arguments but also using analytical ones. We leave as an open question for future investigation the dynamics in the superfluid regime at intermediate and large interactions and whether an interaction-dependent power-law decay of the density is recovered in that regime. Lastly, we have generalized our EoM approach to the BH model on a $D$-dimensional hypercubic lattice both for on-site and long-range two-body losses. 

This paper unraveled important physical results regarding the dissipative quench dynamics of the BH model which can be confirmed experimentally using quantum simulators based on ultra-cold atoms. It also paves the way to the possibility of performing global quench spectroscopy for open quantum lattice models. The latter question can be answered at least for the BH model confined in the SF phase. It requires to use our EoM approach permitting to compute correlation functions for a dissipation- and interaction-quench and then to deduce the quench spectral function by performing a space-time Fourier transform ~\cite{despres2025,jdespres2019,villa2020}.

\section*{Acknowledgments}
We acknowledge funding from the European Research Council (ERC) under the European Union's Horizon 2020 research and innovation programme (Grant agreement No. 101002955 -- CONQUER),  from the Région Île-de-France in the framework of DIM QuanTiP and from the ANR project LOQUST ANR-23-CE47-0006-02. The computations were performed on the LPTMS computer cluster. J. Despres would like to thank M. Cheneau for useful discussions and Z. Qin for the technical support. 

\appendix

\section{Properties of the EoMs for on-site two-body losses}
\label{appendix_prop}

We provide here details about the main properties of the set of EoMs given at Eq.~\eqref{EoMs}. \\

\paragraph{Special case $\gamma = 0$} If $\gamma = 0$ then no quench is performed and the unitary real-time evolution operator $\exp(-i \hat{H}t)$ gives rise to a global phase term $\exp(-i E_{\mathrm{gs}}t)$ for the time-evolved quantum state where $E_{\mathrm{gs}}$ represents the ground state energy of the Hamiltonian $\hat{H}$. Consequently, the time-dependent expectation value of any observable $\hat{O}$ denoted by $\langle \hat{O} \rangle_t$ remains in its initial value, i.e. $\langle \hat{O} \rangle_t = \langle \hat{O} \rangle_0$. This implies that the EoMs are strictly equal to zero at any time $t$, i.e. $\mathrm{d}\langle \hat{O} \rangle_t / \mathrm{d}t = 0$. The latter property is satisfied by the set of EoMs at Eq.~\eqref{EoMs}. Indeed, for $\gamma = 0$, the EoMs reduce to:

\begin{align}
& \frac{\mathrm{d}}{\mathrm{d}t} F_k(t) = -2i\mathcal{A}_{k}(t)F_k(t) - i\mathcal{B}_{k}(t)(1+2G_k(t));\\
& \frac{\mathrm{d}}{\mathrm{d}t} G_k(t) =  -2\mathcal{B}_{k}(t)\operatorname{Im}(F_k(t)); \\
& \frac{\mathrm{d}}{\mathrm{d}t} F_0(t) =  0, \\
& \frac{\mathrm{d}}{\mathrm{d}t} G_0(t) = 0. 
\end{align}

\noindent
At $t = 0$, we know the theoretical expression of $F_k(0)$ and $G_k(0)$ given by:

\begin{equation}
G_k(0) = \frac{1}{2} \left(\frac{\mathcal{A}_{k}}{\mathcal{E}_{k}} - 1 \right),~~~~F_k(0) = -\frac{\mathcal{B}_{k}}{2\mathcal{E}_{k}}.
\end{equation}

\noindent
This leads to $\mathrm{d}F_k(0)/\mathrm{d}t = \mathrm{d}G_k(0)/\mathrm{d}t = 0$. Hence, $F_k(\delta t) = F_k(0)$ and $G_k(\delta t) = G_k(0)$. Then, by calculating 
$\mathrm{d}F_k(\delta t)/\mathrm{d}t$ and $\mathrm{d}G_k(\delta t)/\mathrm{d}t$ and using the previous relations as well as $\mathcal{A}_{k}(\delta t) = \mathcal{A}_{k}(0)$ and $\mathcal{B}_{k}(\delta t) = \mathcal{B}_{k}(0)$, we end up with $\mathrm{d}F_k(\delta t)/\mathrm{d}t = \mathrm{d}G_k(\delta t)/\mathrm{d}t = 0$. Finally, by repeating the previous procedure for each time-step, we find that the EoMs are always equal to zero. \\

\paragraph{Non-interacting case} Another special case corresponds to the non-interacting case where $U = 0$. This implies that $G_k(0) = F_k(0) = F_0(0) = \mathcal{B}_{k}(t) = 0$. According to the EoMs at Eq.~\eqref{EoMs}, we deduce that $G_k(t) = F_k(t) = F_0(t) = 0$. Hence, we get $N(t) = G_0(t)$ and the following EoM associated to $N(t)$:

\begin{align}
& \frac{\mathrm{d}}{\mathrm{d}t} N(t) = -\frac{2\gamma}{L}N(t)(N(t)-1).
\end{align}

\noindent
The latter admits an analytical solution given by:

\begin{equation}
N(t) = \frac{1}{1-e^{-(at+c)}},~a = \frac{2\gamma}{L},~c = \mathrm{ln}\left(\frac{N(0)}{N(0)-1}\right),
\label{N_U_0}
\end{equation}

\noindent
where $N(0) = \bar{n}L$ denotes the total number of bosonic atoms initially present on the lattice.
Therefore, for the condensate density $n(t)$ and the total number of bosonic particles $N(t)$ in the thermodynamic limit, i.e. $L \rightarrow +\infty$, with a large initial number of bosons on the lattice, i.e. $N(0) \rightarrow +\infty$, as well as for small observation times, we get~\cite{liu2022,liu2024,bouchoule2020}:

\begin{equation}
n(t) = \frac{\bar{n}}{1 + 2\gamma \bar{n}t},~~~~~~~N(t) = \frac{N(0)}{1 + 2\gamma \bar{n}t}.
\label{eq_density_U_0}
\end{equation}

\begin{figure}
\includegraphics[scale = 0.4]{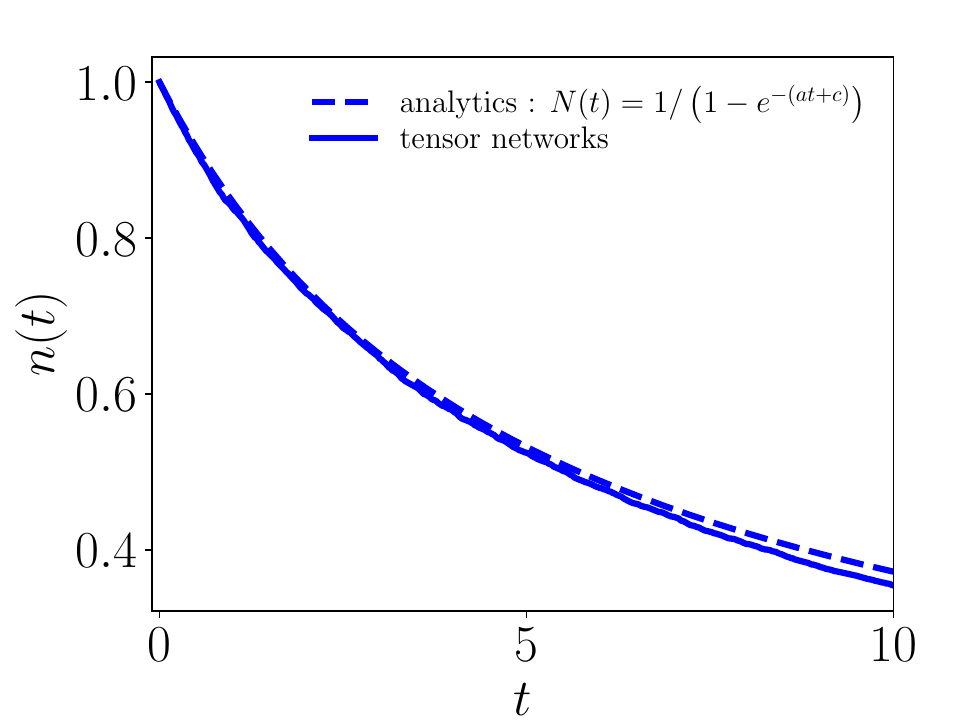}
\caption{Bosonic density $n(t)$ as a function of time $t$ for a sudden global quench on the dissipation strength 
from $\gamma = 0$ to $\gamma = 0.1$ of the non-interacting ($U = 0$) BH chain in the SF-mean-field regime submitted to on-site two-body losses. The solid green line represents the numerical result obtained from the quantum jump method using tensor networks whereas the dashed blue line corresponds to the theoretical solution given at Eq.~\eqref{N_U_0}. The parameters are: $N(0) = L = 12$, $J=1$.}
\label{non_interacting}
\end{figure}

\noindent
The latter clearly shows a long-term decay as $t^{-1}$. Note that even if the previous theoretical result seems to be admitted by the community, it has not been clearly characterized. On Fig.~\ref{non_interacting}, we benchmark our theoretical solution provided at Eq.~\eqref{N_U_0} for the non-interacting case and a finite system size with a tensor networks numerical simulation based on the quantum jump method, see Appendix~\ref{appendix_quantum_jump}. \\

\paragraph{Pseudo-closed equation associated to the density} We now move on to the pseudo-closed equation associated to the density. The latter consists in finding the simplest first-order differential equation associated to the density using the set of EoMs and more restrictive approximations. In what follows, we present a second
theoretical approach to unveil the pseudo-closed equation of the density $n(t)$. The latter is based on the Lindblad master equation applied to the total occupation
number operator $\hat{N}$. The pseudo-closed equation associated to $n(t)$ is deduced from Eq.~\eqref{lindblad_n} where we summed over the momentum. We find:

\begin{align}
& \frac{\mathrm{d}}{\mathrm{d}t}N(t) = i \left \langle \left[\hat{H}, \hat{N} \right] \right \rangle_t + \frac{1}{2}\sum_R \left( \left \langle \hat{L}_R^{\dag}\left[\hat{N}, \hat{L}_R \right] \right \rangle_t + \mathrm{h.c} \right),
\end{align}

\noindent
where the two previous commutators are given by $[\hat{N}, \hat{L}_R] = -2\hat{L}_R$ and $[\hat{H},\hat{N}] = 0$ with $\hat{H}$
defined at Eq.~\eqref{bhm}. We end up with:

\begin{align}
& \frac{\mathrm{d}}{\mathrm{d}t}N(t) = -2 \gamma \sum_R \langle \hat{b}^{\dag}_R \hat{b}^{\dag}_R \hat{b}_R \hat{b}_R \rangle_t.
\label{eq_lme}
\end{align}

\noindent
By going into the reciprocal space and by performing a decoupling of the condensate mode $k = 0$ from the finite modes $k \neq 0$ while using the mean-field approximation as well as $\langle \hat{n}_0^2\rangle_t = \langle \hat{n}_0 \rangle_t^2$, the EoM associated to $n(t)$ is given by:

\begin{align}
& \frac{\mathrm{d}}{\mathrm{d}t}n(t) = - \frac{4\gamma}{L^2}\sum_{q \neq 0} \left[ \operatorname{Re}\left(F_0(t)^* F_q(t)\right) + 2G_0(t) G_q(t) \right] \nonumber  \\
& - \frac{2\gamma}{L^2}G_0(t)(G_0(t)-1).
\end{align} 

\noindent
The last step consists in using the same approximations as those considered in the first approach which are $(1)$ the thermodynamic limit, i.e. 
$L \rightarrow +\infty$ $(2)$ the properties $\operatorname{Re}(F_0(t)) \gg |\operatorname{Im}(F_0(t))|$ and $\vert \sum_{q\neq0}\operatorname{Re}(F_q(t)) \vert \gg \vert \sum_{q\neq0} \operatorname{Im}(F_q(t)) \vert$ leading to $\operatorname{Re}(F_0(t)) = G_0(t)$ $(3)$ the standard mean-field approximation $N(t) \gg \sum_{q \neq 0}G_q(t)$ after having
replaced $\operatorname{Re}(F_0(t))$ by $G_0(t)$. Finally, we end up with the exact same theoretical expression for the pseudo-closed equation of the density given at Eq.~\eqref{eq_pce}. 

\section{Generalization of the EoMs to the 2D square and the $D$-dimensional hypercubic lattices}
\label{2d_eoms}
We now turn to a generalization of the set of EoMs valid for the lattice chain to the 2D square and $D$-dimensional hypercubic lattices.
We start by discussing the dissipative quench dynamics of the 2D BH model on a square lattice initially confined in the SF-mean-field regime for on-site two-body losses. The corresponding Hamiltonian $\hat{H}$ is given by:  

\begin{equation}
\hat{H} = \frac{1}{2} \sum_{\mathbf{k} \neq \mathbf{0}} \mathcal{A}_{\mathbf{k}} \left(\hat{b}^{\dag}_{\mathbf{k}} \hat{b}_{\mathbf{k}} + \hat{b}_{-\mathbf{k}}\hat{b}^{\dag}_{-\mathbf{k}}\right) + \mathcal{B}_{\mathbf{k}} \left(\hat{b}^{\dag}_{\mathbf{k}} \hat{b}^{\dag}_{-\mathbf{k}} + \hat{b}_{\mathbf{k}} \hat{b}_{-\mathbf{k}} \right),
\label{eq_2D_H}
\end{equation}

\noindent
where the coefficients $\mathcal{A}_{\mathbf{k}}$ and $\mathcal{B}_{\mathbf{k}}$ read as:
\begin{align}
& \mathcal{A}_{\mathbf{k}} = 4J\left[\sin^2\left(\frac{\mathbf{k} \cdot \mathbf{x}}{2} \right) + \sin^2\left(\frac{\mathbf{k} \cdot \mathbf{y}}{2} \right)\right] + U\bar{n} \label{eq_2D_ak}; \\
& \mathcal{B}_{\mathbf{k}} = U\bar{n}.
\end{align}

\noindent
The 2D vectors $\mathbf{x}$ and $\mathbf{y}$ and $\mathbf{k}$ are defined as $\mathbf{x} = (a_x, 0)^{T}$, $\mathbf{y} = (0, a_y)^{T}$ and $\mathbf{k} = (k_x, k_y)^{T}$ with $a_x$ and $a_y$ being the lattice spacing along the $x$- and $y$-direction respectively. The latter fulfill $a_x = a_y = 1$. The diagonalized form and the corresponding low-lying excitation spectrum $\mathcal{E}_{\mathbf{k}}$ associated to $\hat{H}$ defined at Eq.~\eqref{eq_2D_H} are given by:
\begin{equation}
\hat{H} = \sum_{\mathbf{k} \neq \mathbf{0}} \mathcal{E}_{\mathbf{k}} \hat{\beta}^{\dag}_{\mathbf{k}} \hat{\beta}_{\mathbf{k}},~~~~~~~~\mathcal{E}_{\mathbf{k}} = \sqrt{\mathcal{A}_{\mathbf{k}}^2 - \mathcal{B}_{\mathbf{k}}^2}.
\end{equation}

\noindent
Employing the same mathematical procedure used for the BH chain, the set of EoMs presented at Eq.~\eqref{eq_2D_EoMs} is found where the initial conditions are given by:
\begin{subequations}
\label{init_2d}
\begin{align}
G_{\mathbf{0}}(0) =&~N_{\mathbf{0}} = N - \sum_{\mathbf{k} \neq \mathbf{0}} G_{\mathbf{k}}(0);\\ 
F_{\mathbf{0}}(0) =&~ \theta(U)N_{\mathbf{0}}; \label{eq_2D_ic} \\
G_{\mathbf{k}}(0) =&~ \frac{1}{2} \left(\frac{\mathcal{A}_{\mathbf{k}}(0)}{\mathcal{E}_{\mathbf{k}}(0)} - 1 \right);\\
F_{\mathbf{k}}(0) =&~ -\frac{\mathcal{B}_{\mathbf{k}}(0)}{2\mathcal{E}_{\mathbf{k}}(0)}, \label{eq_2D_ic_2}
\end{align}
\end{subequations}
and the time-dependent low-lying excitation spectrum and coefficients are defined as:
\begin{subequations}
\label{eq_2D_ic_3}
\begin{align}
\mathcal{E}_{\mathbf{k}}(t) =&~ \sqrt{\mathcal{A}_{\mathbf{k}}(t)^2 - \mathcal{B}_{\mathbf{k}}(t)^2}; \\
\mathcal{A}_{\mathbf{k}}(t) =&~ 4J\left[\sin^2\left(\frac{\mathbf{k} \cdot \mathbf{x}}{2} \right) + \sin^2\left(\frac{\mathbf{k} \cdot \mathbf{y}}{2} \right)\right] + \mathcal{B}_{\mathbf{k}}(t); \\
\mathcal{B}_{\mathbf{k}}(t) =&~\frac{U}{L^2}\sum_{\mathbf{q}} G_{\mathbf{q}}(t).
\end{align}
\end{subequations}

In Appendix~\ref{appendix_2D_prop}, we provide further details on the properties of the set of EoMs as well as the pseudo-closed equation associated to the density. According to the previous theoretical results regarding the dissipative quench dynamics induced by on-site or long-range two-body losses for the BH model on a 1D lattice and a 2D square lattice, we can easily generalise our EoM approach to higher dimensional lattices, see Appendix~\ref{appendix_hypercubic}.

\begin{widetext}
\begin{subequations}
\label{eq_2D_EoMs}
\begin{align}
& \frac{\mathrm{d}}{\mathrm{d}t} G_{\mathbf{0}}(t) = -\frac{\gamma}{L^2}\sum_{\mathbf{q} \neq \mathbf{0}
}\left(F_{\mathbf{0}}(t)F_{\mathbf{q}}(t)^* + \mathrm{h.c}\right) - \frac{4\gamma}{L^2}G_{\mathbf{0}}(t)\sum_{\mathbf{q} \neq \mathbf{0}}G_{\mathbf{q}}(t) - \frac{2\gamma}{L^2}G_{\mathbf{0}}(t)\left(G_{\mathbf{0}}(t)-1\right); \\
& \frac{\mathrm{d}}{\mathrm{d}t} F_{\mathbf{0}}(t) =  -\frac{\gamma}{L^2}\left(2G_{\mathbf{0}}(t)-3\right)F_{\mathbf{0}}(t) - \frac{4\gamma}{L^2}F_{\mathbf{0}}(t)\sum_{\mathbf{q}\neq \mathbf{0}}G_{\mathbf{q}}(t) - \frac{\gamma}{L^2}(2G_{\mathbf{0}}(t)+1)\sum_{\mathbf{q}\neq \mathbf{0}} F_{\mathbf{q}}(t); \\
& \frac{\mathrm{d}}{\mathrm{d}t} G_{\mathbf{k}}(t) =  -2\mathcal{B}_{\mathbf{k}}(t)\operatorname{Im}(F_{\mathbf{k}}(t)) - \frac{\gamma}{L^2}
\left(F_{\mathbf{0}}(t)F_{\mathbf{k}}(t)^* + \mathrm{h.c} \right) -\frac{4\gamma}{L^2}G_{\mathbf{0}}(t)G_{\mathbf{k}}(t),~~~~ \forall \mathbf{k} \neq \mathbf{0}; \\
& \frac{\mathrm{d}}{\mathrm{d}t} F_{\mathbf{k}}(t) = -\left[2i\mathcal{A}_{\mathbf{k}}(t)+\frac{4\gamma}{L^2}G_{\mathbf{0}}(t)\right]F_{\mathbf{k}}(t) - \left[i\mathcal{B}_{\mathbf{k}}(t)+ \frac{\gamma}{L^2}F_{\mathbf{0}}(t)\right](2G_{\mathbf{k}}(t)+1),~~~~ \forall \mathbf{k} \neq \mathbf{0}.
\end{align}
\end{subequations}
\end{widetext}

\section{Properties of the EoMs for long-range two-body losses}
\label{appendix_non_local}
In what follows, we briefly describe the properties of the set of EoMs at Eq.~\eqref{EoMs_nl} associated to the 1D BH model with long-range two-body losses.
For the non-interacting case, i.e. $U = 0$, the time-dependent total occupation number $N(t)$ obeys the following first-order differential equation:

\begin{align}
& \frac{\mathrm{d}}{\mathrm{d}t} N(t) = -2\mathcal{G}_{0} N(t)(N(t)-1),
\end{align}

\noindent
which can be solved analytically and leading to the expression:

\begin{equation}
N(t) = \frac{1}{1-e^{-(2\mathcal{G}_{0}t+c)}},~~~~~c = \mathrm{ln}\left(\frac{N(0)}{N(0)-1}\right).
\end{equation}
 
\noindent
In the thermodynamic limit, i.e. $L\rightarrow +\infty$, while considering a large initial number of bosons on the lattice, i.e. 
$N(0) \rightarrow + \infty$, as well as small observation times, $n(t)$ and $N(t)$ are thus given by: 
 
\begin{equation}
n(t) = \frac{\bar{n}}{1 + 2\tilde{\gamma}(0)\bar{n}t},~~~~~~N(t) = \frac{N(0)}{1 + 2\tilde{\gamma}(0)\bar{n}t},
\label{density_nl_non_int}
\end{equation}

\noindent
with $\tilde{\gamma}(q) = L\mathcal{G}_q$. Concerning the pseudo-closed equation of $n(t)$ for long-range two-body losses, the latter is deduced using the same approximations than those we considered to unveil the one for on-site two-body losses at Eq.~\eqref{eq_pce} and reads as:

\begin{align}
& \frac{\mathrm{d}}{\mathrm{d}t} n(t) = -4n(t)\frac{1}{L}\sum_{q\neq0}\tilde{\gamma}(q)\left[G_q(t) + \operatorname{Re}(F_q(t))\right]
\nonumber \\
& -2\tilde{\gamma}(0) n(t)^2.
\label{density_nl_pce}
\end{align}

\noindent
We verified that the latter reproduces accurately the time-dependent density profile computed from the coupled EoMs given at Eq.~\eqref{EoMs_nl}.

\section{Properties of the EoMs for the 2D square lattice}
\label{appendix_2D_prop}

We provide here further information about the properties of the EoMs valid in the 2D case as well as the associated pseudo-closed equation for the density. 
According to Eq.~\eqref{eq_2D_EoMs}, we get in the non-interacting case, i.e. $U = 0$:

\begin{align}
& \frac{\mathrm{d}}{\mathrm{d}t} N(t) = -\frac{2\gamma}{L^2}N(t)(N(t)-1),
\end{align}

\noindent
which yields for the total number of bosons $N(t)$:

\begin{equation}
N(t) = \frac{1}{1-e^{-(at+c)}}, a = \frac{2\gamma}{L^2}, c = \mathrm{ln}\left(\frac{N(0)}{N(0)-1}\right).
\label{eq_density_U_0_2D_first}
\end{equation}

\noindent
Finally, in the limit $L,N(0) \rightarrow + \infty$ while considering short observation times, the condensate density $n(t)$
and the total occupation number $N(t)$ are given, as expected, by: 

\begin{equation}
n(t) = \frac{\bar{n}}{1 + 2\gamma \bar{n}t},~~~~~~ N(t) = \frac{N(0)}{1 + 2\gamma \bar{n}t},
\label{eq_density_U_0_2D}
\end{equation}

\noindent
where $n(0) = \bar{n} = N(0)/L^2$ refers to the initial density (or filling) of the 2D square lattice. Regarding the pseudo-closed equation of the density $n(t)$, the latter can be easily generalized to the 2D case and reads as:

\begin{align}
& \frac{\mathrm{d}}{\mathrm{d}t} n(t) = -2\gamma n(t)^2  -4\gamma n(t)\frac{1}{L^2}\sum_{\mathbf{q} \neq \mathbf{0}}\left[G_{\mathbf{q}}(t) + \operatorname{Re}(F_{\mathbf{q}}(t))\right].
\label{eq_2D_pce}
\end{align}

\noindent
Similarly to the 1D case, we verified that the latter reproduces the time-dependent density profile computed from the coupled EoMs at Eq.~\eqref{eq_2D_EoMs}. 

\section{Properties of the density for a $D$-dimensional hypercube}
\label{appendix_hypercubic}

Let us consider from now a $D$-dimensional hypercubic lattice. The Hamiltonian $\hat{H}$ has the same expression as the one presented at Eq.~\eqref{eq_2D_H} except that the sum is now performed on the $D$-dimensional wave-vector $\mathbf{k}$
defined as $\mathbf{k} = (k_x, k_y, k_z, ...)^{T}$. Besides, the coefficient $\mathcal{A}_{\mathbf{k}}$ becomes:
\begin{equation}
\mathcal{A}_{\mathbf{k}} = 4J\sum_{i=1}^{D} \sin^2\left(\frac{\mathbf{k} \cdot \mathbf{d}_i}{2}\right) + U\bar{n},
\end{equation}
\noindent
where $\mathbf{d}_1 = (a_x, 0, ..., 0)^{T}$, $\mathbf{d}_2 = (0, a_y, 0, ..., 0)^{T}$, $\mathbf{d}_i = (0, ..., 0, a_i, 0, ..., 0)^{T}$ with $a_x = a_y = a_i = 1$ and $n(0) = \bar{n} = N(0)/L^D$ refers to the initial density (or filling) of the $D$-dimensional hypercube. The initial conditions for the set of EoMs are the same as those presented previously for the 2D case, see Eq.~\eqref{init_2d}. However, Eq.~\eqref{eq_2D_ic_3} has to be slightly modified and becomes:
\begin{align}
\mathcal{A}_{\mathbf{k}}(t) =&~4J\sum_{i=1}^{D} \sin^2\left(\frac{\mathbf{k} \cdot \mathbf{d}_i}{2}\right) + \mathcal{B}_{\mathbf{k}}(t) \label{ak_hypercube}, \\
\mathcal{B}_{\mathbf{k}}(t) =&~ \frac{U}{L^D} \sum_{\mathbf{q}} G_{\mathbf{q}}(t).
\end{align}
Concerning the theoretical expression of the EoMs for the $D$-dimensional hypercube in the case of on-site two-body losses, the latter are the same as those presented at Eq.~\eqref{eq_2D_EoMs} for the 2D square lattice except that the area associated to the 2D square lattice $\mathcal{A} = L^2$ has to be replaced by the volume of the hypercubic lattice $\Omega = L^D$ and the wave-vector $\mathbf{k}$ is now homogeneous to a $D$-dimensional vector. For long-range two-body losses, the EoMs are the same as those presented at Eq.~\eqref{EoMs_nl} for the 1D lattice except that the chain length $L$ is replaced by the hypercube volume $L^D$ and the wave-vector is no longer homogeneous to a scalar but to a $D$-dimensional vector. Furthermore, the quasimomentum-dependent functions become:
\begin{align}
\mathcal{F}_{\mathbf{q}} =&~ 2[\mathcal{G}_{\mathbf{0}}+\mathcal{G}_{\mathbf{q}}];\\
\mathcal{G}_{\mathbf{q}} =&~ \frac{1}{L^{2D}}\sum_{\mathbf{R},\mathbf{R'}}\gamma_{||\mathbf{R}-\mathbf{R'}||}\cos(\mathbf{q}(\mathbf{R}-\mathbf{R'})); \\
\mathcal{H}_{\mathbf{q}} =&~ \frac{1}{L^{2D}}\sum_{\mathbf{R},\mathbf{R'}}\gamma_{||\mathbf{R}-\mathbf{R'}|}i\sin(\mathbf{q}(\mathbf{R}-\mathbf{R'})),
\end{align}
where the long-range dissipation strength $\gamma_{||\mathbf{R}-\mathbf{R'}||}$ reads as:
\begin{equation}
\gamma_{||\mathbf{R}-\mathbf{R'}||} = \frac{\Gamma}{(1+||\mathbf{R}-\mathbf{R'}||)^{\alpha}},
\end{equation}
with $||...||$ standing for the vector norm. For on-site two-body losses and from our findings related to the 2D case, see Appendix~\ref{appendix_2D_prop}, we can easily generalize the theoretical expression of the density $n(t)$ in the non-interacting limit to the $D$-dimensional hypercubic lattice as well as its pseudo-closed equation in the SF-mean-field regime. The latter are respectively given by:
\begin{align}
& n(t) = \frac{\bar{n}}{1+2 \gamma \bar{n}t};~~~~\bar{n} = n(0) = \frac{N(0)}{L^D},\\
& \frac{\mathrm{d}}{\mathrm{d}t} n(t) = -2\gamma n(t)^2  -4\gamma n(t)\frac{1}{L^D}\sum_{\mathbf{q} \neq \mathbf{0}}\left[G_{\mathbf{q}}(t) + \operatorname{Re}(F_{\mathbf{q}}(t))\right].
\end{align}

\noindent
For long-range two-body losses and from our findings related to the 1D case, see Appendix~\ref{appendix_non_local}, we can generalize the theoretical expression of the density $n(t)$ as well as the total number of bosonic atoms on the lattice $N(t)$ in the non-interacting limit to the $D$-dimensional hypercubic lattice. The density $n(t)$ in the non-interacting case reads as:

\begin{align}
& n(t) = \frac{\bar{n}}{1 + 2\tilde{\gamma}(\mathbf{0})\bar{n}t};~~~~N(t) = \frac{N(0)}{1 + 2\tilde{\gamma}(\mathbf{0})\bar{n}t},
\end{align}

\noindent
with $\bar{n} = N(0)/L^D$ and the pseudo-closed equation associated to $n(t)$ in the SF-mean-field regime is given by:

\begin{align}
& \frac{\mathrm{d}}{\mathrm{d}t} n(t) = -4n(t)\frac{1}{L^D}\sum_{\mathbf{q} \neq \mathbf{0}}\tilde{\gamma}(\mathbf{q})\left[G_{\mathbf{q}}(t) + \operatorname{Re}(F_{\mathbf{q}}(t))\right] \nonumber \\
& -2\tilde{\gamma}(\mathbf{0}) n(t)^2,
\end{align}

\noindent
where the function $\tilde{\gamma}$ is defined as follows:

\begin{equation}
\tilde{\gamma}(\mathbf{q}) = \frac{1}{L^D} \sum_{\mathbf{R},\mathbf{R'}} \gamma_{||\mathbf{R}-\mathbf{R'}||} \cos(\mathbf{q}(\mathbf{R}-\mathbf{R'})).
\end{equation}

\section{Quantum jump method}
\label{appendix_quantum_jump}
We give here further details concerning the quantum jump method permitting to solve numerically the Lindblad master equation to characterize the dissipative dynamics of open quantum lattice models. In what follows, we will work along the lines of Ref.~\cite{daley2014}. This numerical method offers the possibility to unravel the Lindblad master equation via a set of quantum trajectories describing the real time evolution of the open quantum lattice model conditioned to a set of measurement outcomes. More precisely, the real time evolution of the quantum lattice model is described by a stochastic Schr\"odinger equation where at random times a Lindblad jump operator
$\hat{L}_{R}$ is applied to the time-evolved quantum state while in between quantum jumps the real time evolution of the quantum lattice model is non-unitary and driven by a non-Hermitian Hamiltonian. In what follows, we provide a mathematical transcription of the quantum jump protocol for on-site and long-range two-body losses. \\

For on-site two-body losses, the Lindblad master equation reads:
\begin{align}
\frac{\mathrm{d}}{\mathrm{d}t}\hat{\rho} = -i \left[\hat{H},\hat{\rho}\right] + \sum_R \hat{L}_R \hat{\rho} \hat{L}^{\dag}_R - \frac{1}{2} \left \{\hat{L}^{\dag}_R \hat{L}_R, \hat{\rho} \right \},
\end{align}
\noindent
where $\hat{L}_R = \sqrt{\gamma} \hat{b}_R^2$ refers to the Lindblad jump operator acting on the lattice site $R$ and $\hat{H}$
to the Hermitian Hamiltonian of the BH chain, see Eq.~\eqref{bhm}. The latter can be rewritten as:
\begin{equation}
\frac{\mathrm{d}}{\mathrm{d}t}\hat{\rho} = -i \left( \hat{H}_{\mathrm{eff}}\hat{\rho} - \hat{\rho}\hat{H}^{\dag}_{\mathrm{eff}} \right) + \sum_R \hat{L}_R \hat{\rho} \hat{L}^{\dag}_R,
\end{equation}
\noindent
where the effective non-Hermitian Hamiltonian $\hat{H}_{\mathrm{eff}}$ is given by:
\begin{equation}
\hat{H}_{\mathrm{eff}} = \hat{H} - \frac{i}{2} \sum_R \hat{L}^{\dag}_R \hat{L}_R.
\end{equation}
\noindent
Supposing that we have a well-defined initial state $\ket{\Psi(0)}$, the following procedure has to be performed for each time-step $\delta t$: \\

\noindent
Starting from time $t$ with quantum state $\ket{\Psi(t)}$; to deduce the time-evolved quantum state at time $t+\delta t$, we begin by computing the candidate:
 \begin{equation}
 \ket{\Psi^{(1)}(t+\delta t)} = (1-i\hat{H}_{\mathrm{eff}}\delta t) \ket{\Psi(t)}.
 \end{equation}
 \noindent
 The latter quantum state has a norm to the square at first order in $\delta t$ given by:
 
 \begin{equation}
 \langle \Psi^{(1)}(t+\delta t) | \Psi^{(1)}(t+\delta t) \rangle = 1 - \delta p  < 1,
 \end{equation}
 \noindent
 due to the non-hermiticity of $\hat{H}_{\mathrm{eff}}$ where:
 \begin{align}
  & \delta p = \delta t \sum_R \langle \Psi(t)| \hat{L}^{\dag}_R \hat{L}_R | \Psi(t) \rangle = \sum_R \delta p_R,
 \end{align}
  \noindent 
  and $\delta p_R = \delta t \langle \Psi(t)| \hat{L}^{\dag}_R \hat{L}_R | \Psi(t) \rangle$ represents the probability that the quantum jump occurs on the lattice site $R$ for this specific time-step. \\
  
  \noindent
  Then, the quantum state $\ket{\Psi(t)}$ propagates in time stochastically according to the following scheme:\\
  \noindent
  - With a probability $1-\delta p$, we affect the candidate state $\ket{\Psi^{(1)}(t+\delta t)}$ to the quantum state $\ket{\Psi(t+\delta t)}$. Mathematically, it means:
  \begin{equation}
   \ket{\Psi(t+\delta t)} = \frac{\ket{\Psi^{(1)}(t+\delta t)}}{||\ket{\Psi^{(1)}(t+\delta t)}||}.
  \end{equation}
  - With a probability $\delta p$, a quantum jump occurs on a lattice site $R$ and it yields:
  \begin{equation}
  \ket{\Psi(t+\delta t)} = \frac{\hat{L}_R\ket{\Psi(t)}}{||\hat{L}_R\ket{\Psi(t)}||}.
  \end{equation}
  The Lindblad quantum jump is applied on the lattice site $R$ through the Lindblad jump operator $\hat{L}_R$ with a probability given by $\Pi_R = \delta p_R / \delta p$.   
  Finally, to deduce $O(t)$ the expectation value of an observable $\hat{O}$ at time $t$, we need to average over the quantum trajectories:
  \begin{equation}
  O(t) = \frac{1}{N_{\mathrm{traj}}}\sum_{n = 1}^{N_{\mathrm{traj}}} \langle \Psi^{(i)}(t) | \hat{O} | \Psi^{(i)}(t) \rangle.
  \end{equation}

\begin{figure}[t]
\includegraphics[scale = 0.36]{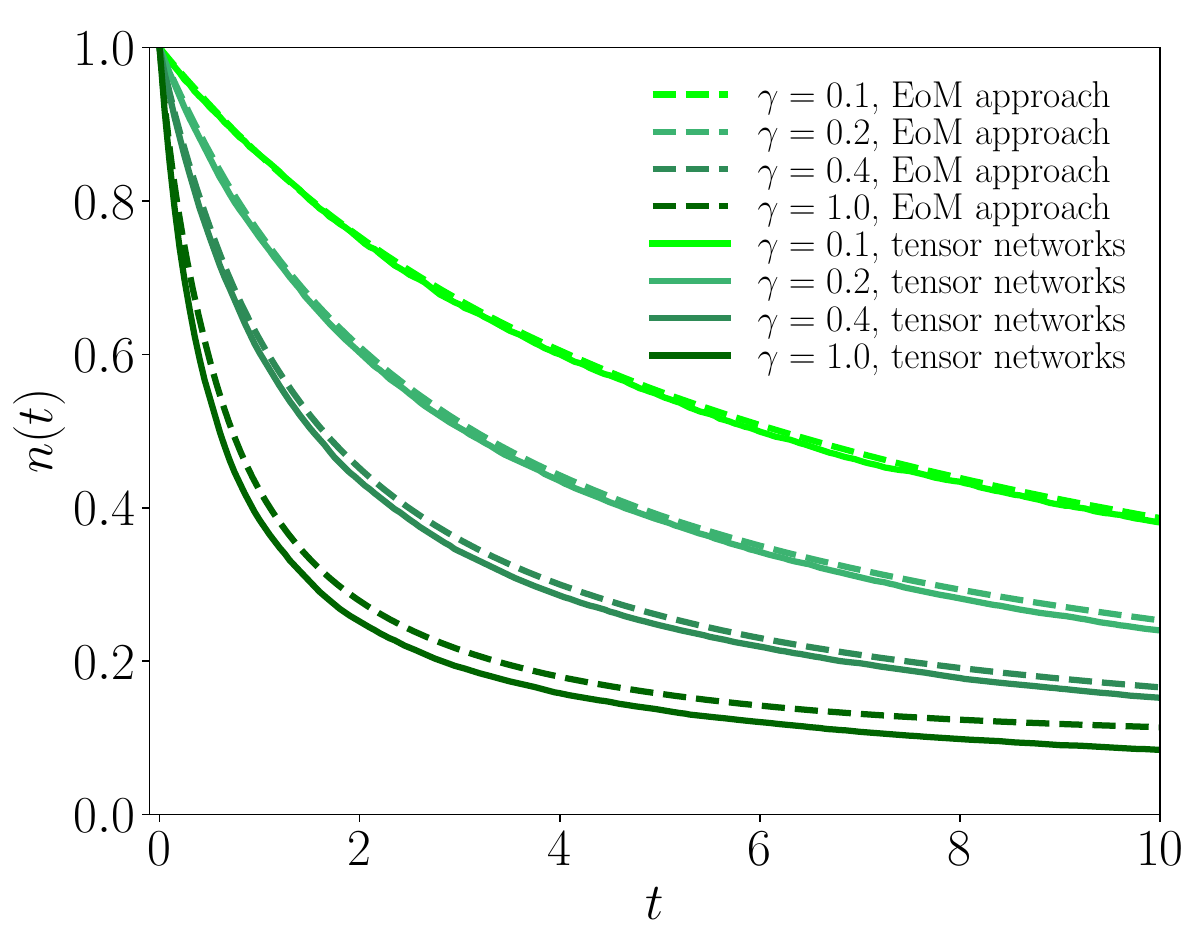}
\caption{Bosonic density $n(t)$ as a function of time $t$ of the BH chain initially confined in the SF-mean-field regime for a sudden global quench on the dissipation strength from $\gamma = 0$ to $\gamma > 0$ at fixed two-body repulsive interaction strength $U$. The solid lines represent numerical results obtained from the quantum jump method using tensor networks whereas the dashed lines correspond to theoretical predictions from the EoM approach given at Eq.~\eqref{EoMs}. The parameters are: $N(0) = L = 18$, $J = 1$, $U = 0.2$.}
\label{gamma_N_18}
\end{figure}

\begin{figure*}[t!]
\includegraphics[scale = 0.44]{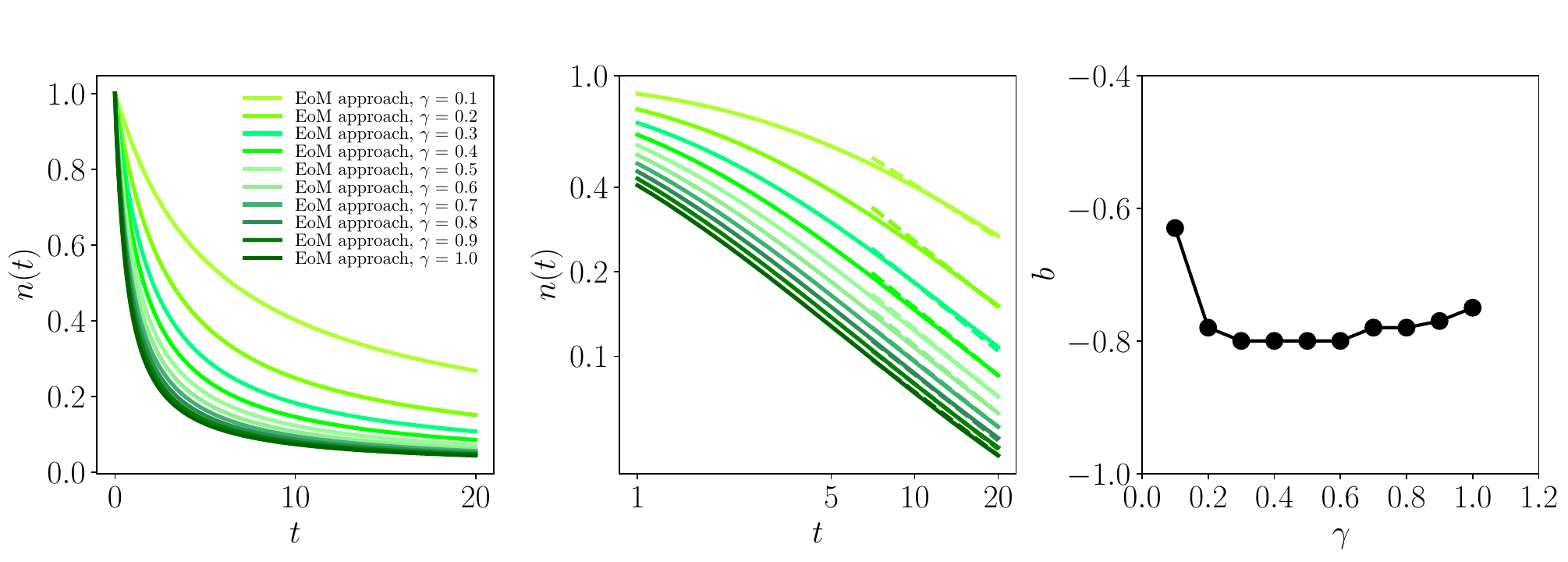}
\caption{Bosonic density $n(t)$ as a function of time $t$ for the BH chain initially confined in the SF-mean-field regime for a sudden global quench on the dissipation strength $\gamma$ from $\gamma = 0$ to $\gamma > 0$. The latter is investigated for various $\gamma$ while fixing the interaction strength $U$. (Left) time-dependent density profile in linear-linear scale (center) in log-log scale. The solid lines represent theoretical results obtained from the EoM approach whereas the dashed lines correspond to fits characterised by an ansatz of the form $n_{\mathrm{fit}}(t) = at^b$.
(Right) power-law exponent $b$ as a function of $\gamma$. The parameters are: $N(0) = L = 100$, $J = 1$, $U = 0.3$.}
\label{fig_fits_U_higher_gamma_varies}
\end{figure*}

We now turn on the case of long-range two-body losses where the generalization is relatively straightforward. The latter requires to consider the previous procedure and to replace the spatial degree of freedom $R$ by $R$ and $R'$ and thus the single sum over $R$ by a double sum over both lattice site indices $R$ and $R'$. The Lindblad jump operator becomes $\hat{L}_{R,R'} = \sqrt{\gamma_{|R-R'|}} \hat{b}_R\hat{b}_{R'}$ and the probability $\delta p$ that a quantum jump occurs is now defined as:
\begin{equation}
\delta p = \delta t \sum_{R,R'} \langle \Psi(t)|\hat{L}^{\dag}_{R,R'} \hat{L}_{R,R'} | \Psi(t) \rangle = \sum_{R,R'} \delta p_{R,R'}.
\end{equation}
\noindent
The probability for the quantum jump to be performed on the specific lattice sites $R$ and $R'$ is now given by $\Pi_{R,R'} = \delta p_{R,R'} / \delta p$.

\section{Additional numerical results}
\label{app:additional}
In this Appendix, we provide additional numerical results regarding the dynamics of our model. \\

\begin{figure}[h]
\includegraphics[scale = 0.46]{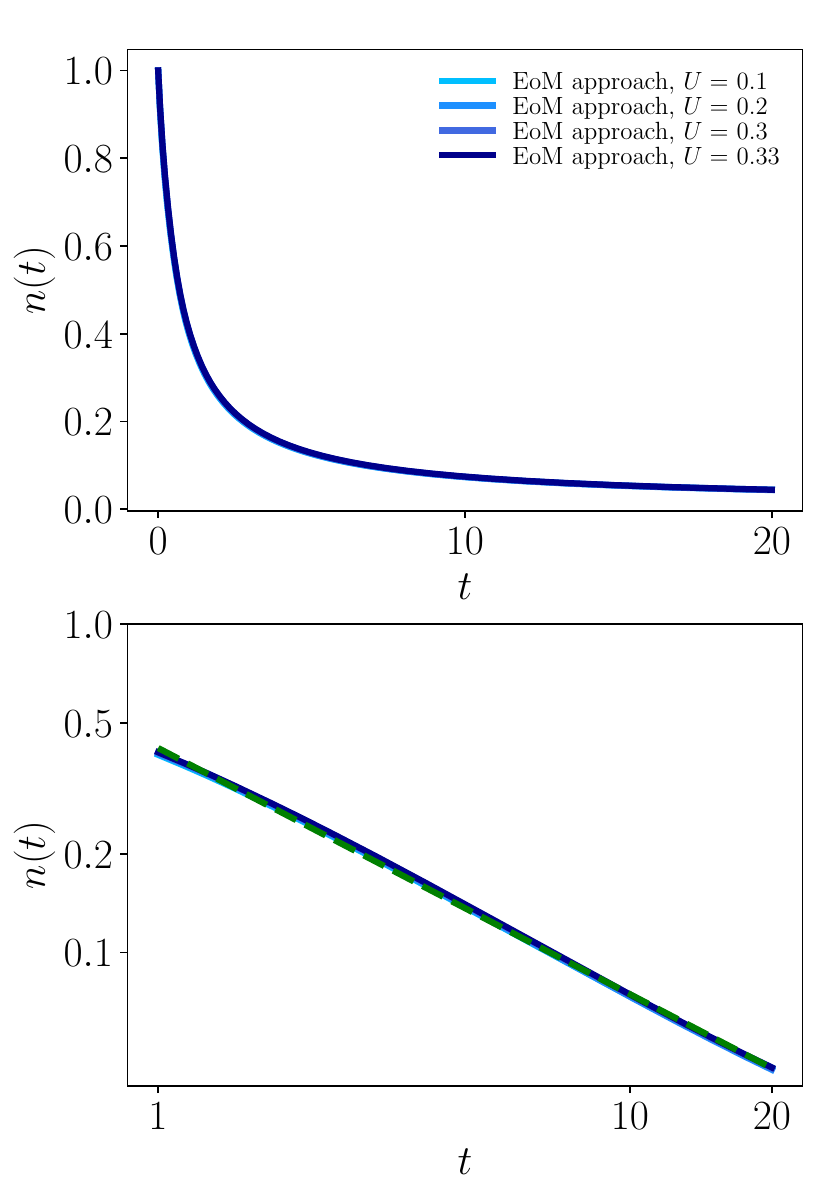}
\caption{Bosonic density $n(t)$ as a function of time $t$ of the BH chain initially confined in the SF-mean-field regime for a sudden global quench on the 
dissipation strength $\gamma$ from $\gamma = 0$ to $\gamma > 0$. The latter is investigated for various interaction strengths $U$ while fixing $\gamma$. (Top) time-dependent density profile in linear-linear scale (bottom) in log-log scale. The solid lines represent theoretical results obtained from the EoM approach whereas the dashed line corresponds to a fit having an ansatz of the form $n_{\mathrm{fit}}(t) = at^b$ where $a = 0.42$ and $b = -0.75$. The parameters are: $N(0) = L = 100$,
$J = 1$, $\gamma = 1$.}
\label{fig_fits_U_varies_gamma_large}
\end{figure}

First, we provide further theoretical and numerical results concerning the time-dependent density $n(t)$ as a function of $\gamma$. 
As depicted on Fig.~\ref{gamma_N_18}, we clearly find a very good agreement between the theoretical predictions deduced from the EoM approach and the numerical results obtained from the quantum jump method using tensor networks for a larger system size, i.e. $L = 18$. 

Then, we present here theoretical results concerning the time-dependent density for a sudden global quench on $\gamma$ at fixed $J$ and $U$ for the BH chain initially confined in the SF-mean-field regime. On Fig.~\ref{fig_fits_U_higher_gamma_varies}, the interaction strength $U$ has a larger value than the one used in the main text, i.e. $U = 0.3$ compared to $U = 0.1$ on Fig.~\ref{fig_fits_U_small_gamma_varies}. Similarly to the case $U = 0.1$; for relatively large values of $\gamma$, i.e. $\gamma \gtrsim 0.2$, the density profiles are shifted while the algebraic decay in time remains unchanged and is characterized by a power-law exponent $b \simeq -0.8$. Note that the previous value of the $\gamma$-independent power-law exponent $b$ is exactly the one found for $U = 0.1$ as depicted on Fig.~\ref{fig_fits_U_small_gamma_varies}. For small dissipation strengths $\gamma$, i.e. $\gamma \lesssim 0.2$, the power-law exponent $b$ does not remain constant as expected, see Fig.~\ref{fig_fits_U_small_gamma_varies}. Indeed, for low $\gamma$, the density $n(t)$ was found to display an algebraic decay characterized by a $U$-dependent power-law exponent $b$. The latter exponent is expected to depend as well on the small value of $\gamma$.

Finally, we present here theoretical results concerning the time-dependent density $n(t)$ for a sudden global quench on $\gamma$ at fixed $J$ and for various values of the interaction strength $U$ for the BH chain initially confined in the SF-mean-field regime. On Fig.~\ref{fig_fits_U_varies_gamma_large}, the dissipation strength
$\gamma$ has a value significantly larger than the one used in the main text, i.e. $\gamma = 1$ compared to $\gamma = 0.1$ on Fig.~\ref{fig_fits_U_varies_gamma_small}. Contrary to the case $\gamma = 0.1$ where the power-law exponent $b$ is $U$-dependent, the density profiles for $\gamma = 1$ are completely dominated by the dissipation leading to the same curve for all the different values of $U$ and thus to the same pre-factor $a$ as well as to the same power-law exponent $b$. Note that the profile in time of the density $n(t)$ has been verified to be independent of the length of the chain $L$, provided that the latter is large enough.  

\section{Time-dependent density characteristics depending on the dimensionality of the lattice}
\label{appendix_dim}
We propose here to unveil by simple theoretical arguments the influence of the lattice dimensionality on the decay in time of the density $n(t)$.
In the following, we focus on the specific case of small interactions and on-site dissipation strengths denoted by $U$ and $\gamma$ respectively. \\

For a 1D lattice, the time-dependent density is found to decrease algebraically with an interaction-dependent power-law exponent, see Fig.~\ref{fig_fits_U_varies_gamma_small}. For the 2D square lattice, the density profile in time is drastically different, namely it displays a very similar decay in time independently of $U$ and is fully characterized by the non-interacting case, i.e. $U = 0$, as depicted on Fig.~\ref{fig_fits_U_varies_gamma_small_2D}. This dependence on the lattice dimensionality can be explained analytically by analyzing the pseudo-closed equation associated to the time-dependent density as well as the corresponding set of EoMs. According to the pseudo-closed equation for $n(t)$ in the 2D case at Eq.~\eqref{eq_2D_pce}, the density being $U$-independent implies that the quadratic correlators $G_{\mathbf{k}}(t)$ and $\operatorname{Re}(F_{\mathbf{k}}(t))$ are also independent of the interaction strength $U$. From the corresponding EoMs at Eq.~\eqref{eq_2D_EoMs}, the dependence on $U$ is contained within the functions $\mathcal{A}_{\mathbf{k}}(t)$ and $\mathcal{B}_{\mathbf{k}}(t)$. Concerning the function $\mathcal{A}_{\mathbf{k}}(t)$, the latter has an additional tight-binding dispersion relation compared to the 1D case due to the second component of the wave-vector associated to the transverse $y$-direction, i.e. $4J\sin^2(\mathbf{k}\cdot \mathbf{y}/2)$. Consequently, $\mathcal{A}_{\mathbf{k}}$ is much larger in the 2D case and the $U$-dependent term within the latter becomes negligible with respect to the sum of the two tight-binding dispersion relations. Then, only the function $\mathcal{B}_{\mathbf{k}}(t)$ involves a possible dependence on $U$. It immediately follows that $\operatorname{Re}(F_{\mathbf{k}}(t))$ is $U$-independent since the latter does not depend on $\mathcal{B}_{\mathbf{k}}(t)$ according to its EoM at Eq.~\eqref{eq_2D_EoMs}. 
We now move on to the EoM associated to $G_{\mathbf{k}}(t)$ where the dependence on $U$ is contained in the function $\mathcal{B}_{\mathbf{k}}(t)$ of the first term. This first term also depends on the correlator $\operatorname{Im}(F_{\mathbf{k}}(t))$ which is negligible with respect to the second and third terms and can thus be disregarded. Finally, we have shown with simple theoretical arguments that neither the correlator $G_{\mathbf{k}}(t)$ nor $\operatorname{Re}(F_{\mathbf{k}}(t))$ depends on the interaction strength $U$ for the 2D BH model on a square lattice. This implies that $n(t)$ is $U$-independent according to its pseudo-closed
equation. \\

Note that a very similar analysis can be performed in order to generalise the previous statement to a 3D cubic lattice up to a $D$-dimensional hypercubic lattice. Indeed, the arguments are exactly the same than those presented previously. We can also add that the argument regarding the function $\mathcal{A}_{\mathbf{k}}$ being larger than its version in the 1D case is even stronger for a hypercube of dimension $D$, see Eq.~\eqref{ak_hypercube}. Indeed, for a $D$-dimensional hypercube, the full tight-binding dispersion relation $\epsilon_{\mathbf{k}}$ contained within $\mathcal{A}_{\mathbf{k}}$ is given by: 
\begin{equation}
\epsilon_{\mathbf{k}} = 4J \sum_{i=1}^{D} \sin^2\left(\frac{\mathbf{k}.\mathbf{d}_i}{2}\right).
\end{equation}
As a consequence, the energies of the dispersion relation $\epsilon_{\mathbf{k}}$ for the hypercube of dimension $D$ will be much larger than those for the lattice
chain implying $D = 1$. To summarize, from the 2D square lattice up to the $D$-dimensional hypercube, the density $n(t)$ will not be $U$-dependent. 

\bibliography{biblioJD,biblioLSP,biblioLV,biblioJDv2,biblioLSPv2,biblioLVv2}

\end{document}